\documentclass[conference]{IEEEtran}
\IEEEoverridecommandlockouts
\usepackage{cite}
\usepackage{amsmath,amssymb,amsfonts}
\usepackage{algorithmic}
\usepackage{graphicx}
\usepackage{textcomp}
\usepackage{xcolor}
\usepackage{amsthm}
\usepackage{leftindex}
\def\BibTeX{{\rm B\kern-.05em{\sc i\kern-.025em b}\kern-.08em
    T\kern-.1667em\lower.7ex\hbox{E}\kern-.125emX}}

\usepackage{csquotes}
\newcommand{\ctl}{{\sf CTL*}}
\newcommand{\ltl}{{\sf LTL}}
\newcommand{\hypltl}{{\sf HyperLTL}}
\newcommand{\hypctl}{{\sf HyperCTL*}}
\newcommand{\timhyp}{timing hyperproperties}

\newcommand{\mtl}{{\sf MTL}}

\newcommand{\mitl}{{\sf MITL}}
\newcommand{\hypmitl}{{\sf HyperMITL}}
\newcommand{\QPTL}{{\sf QPTL}}
\newcommand{\hyptidyctl}{{\sf HyperTidy CTL*}}

\newtheorem{theorem}{Theorem}
\newtheorem{lemma}[theorem]{Lemma}
\newtheorem{corollary}[theorem]{Corollary}
\theoremstyle{definition}

\theoremstyle{remark}
\newtheorem{remark}[theorem]{Remark}
\newtheorem{example}[theorem]{Example}

\newcommand{\ndc}[1] {\textcolor{teal}{ND: #1}}
\newcommand{\mvc}[1] {\textcolor{red!60!black}{MV: #1}}

\newcommand{\rmv}[1] {}
\newcommand{\myparagraph}[1] {\vspace*{0.15cm} \noindent {\bf #1.}}
\newcommand{\s}[1] {\mathsf{#1}}

\newcommand{\arxiv}{}

\newcommand{\nat}{\mathbb{N}}
\newcommand{\reals}{\mathbb{R}}
\newcommand{\nnreals}{\reals_{\geq 0}}
\newcommand{\posreals}{\reals_{> 0}}
\newcommand{\powerset}[1]{2^{#1}}
\newcommand{\set}[1] {\{#1\}}
\newcommand{\setpred}[2] {\{#1\: |\: #2\}}
\newcommand{\cO} {{\cal O}}
\newcommand{\cT}{{\cal T}}
\newcommand{\cB}{{\cal B}}

\newcommand{\props}{P}

\newcommand{\len}[1]{|#1|}

\newcommand{\opb}{(\!|}
\newcommand{\cpb}{|\!)}

\newcommand{\rend}[1]{R(#1)}
\newcommand{\lend}[1]{L(#1)}
\newcommand{\is}{interval sequence}

\newcommand{\cons}{\psi}
\newcommand{\conset}[1]{\Phi(#1)}
\newcommand{\clockval}{\mu}
\newcommand{\ta}{\mathcal{A}}
\newcommand{\st}{V}
\newcommand{\startst}{V_0}
\newcommand{\stlab}{\alpha}
\newcommand{\clocks}{X}
\newcommand{\clockcons}{\beta}
\newcommand{\edges}{E}
\newcommand{\final}{V_F}
\newcommand{\taext}{\ta = (\st, \startst, \stlab, \clocks, \clockcons, \edges, \final)}
\newcommand{\run}{\rho}

\newcommand{\flowrun}{f}

\newcommand{\guard}{\psi}
\newcommand{\guardset}{\Psi}

\newcommand{\mso}{\text{MSO}(<, +1)}

\newcommand{\MP}{\textsf{MP}}
\newcommand{\MPP}{\textsf{P}}
\newcommand{\tdomain}{\mathbb{T}}
\newcommand{\msof}{\varphi}
\newcommand{\flow}{f}
\newcommand{\flowg}{g}
\newcommand{\freevar}{\s{Vars}}
\newcommand{\inter}{I}

\newcommand{\hcmtl}{\textsf{HCMTL*}}
\newcommand{\hcmtlf}{\varphi}
\newcommand{\pvar}{\mathcal{V}}
\newcommand{\Tez}{T^-}
\newcommand{\Teo}{T^+}
\newcommand{\rz}[1]{#1^-}
\newcommand{\ro}[1]{#1^+}
\newcommand{\flows}[1]{\mathsf{exec}(#1)}
\newcommand{\flowsn}[1]{\mathsf{exec}_N(#1)}
\newcommand{\penv}{\Pi}
\newcommand{\dagg}{\dagger}
\newcommand{\hcmtlmodels}{\penv, t, \dagg \models_{\ta}}

\newcommand{\hcmtlmodelsarg}[3]{#1,#2,#3\models_{\ta}}
\newcommand{\hcmtlmodelsbdd}{\penv, t, \dagg \models^N_{\ta}}
\newcommand{\hcmtlmodelsargbdd}[3]{#1,#2,#3\models^N_{\ta}}
\newcommand{\emptypenv}{\{\}}
\newcommand{\exec}{execution}

\newcommand{\runtoflow}[1]{F_{#1}}

\newcommand{\fpi}[1]{f^{#1}}
\newcommand{\tphi}[1]{\leftindex^\ta T^{#1}}

\newcommand{\event}{\sigma}
\newcommand{\evtime}{t}
\newcommand{\ptta}{\mathcal{B}}
\newcommand{\ptst}{S}
\newcommand{\ptstartst}{s_0}
\newcommand{\ptclocks}{X}
\newcommand{\ptedges}{\Delta}
\newcommand{\ptfinal}{F}
\newcommand{\pttaext}{\ptta = (\powerset{\props}, \ptst, \ptstartst, \ptclocks, \ptedges, \ptfinal)}
\newcommand{\ptrun}{\eta}
\newcommand{\ptruns}[1]{\mathsf{exec}^{pt}(#1)}
\newcommand{\ptpenv}{\Gamma}
\newcommand{\pthcmtlmodels}{\pthcmtlargta{\ptpenv}{t}{\dagg}{\ptta}}

\newcommand{\xsing}{x_{sing}}
\newcommand{\enc}{\chi}
\newcommand{\cphi}[1]{{#1}^{\mathsf{ib}}}
\newcommand{\taenc}[1]{#1_{\mathsf{ib}}}
\newcommand{\marker}{\#}
\newcommand{\hcmtlargta}[4]{{#1,#2,#3\models_{#4}}}
\newcommand{\pthcmtlargta}[4]{{#1,#2,#3\leftindex^{pt}{\models}_{#4}}}

\newcommand{\msotrees}{S2S}

\title{Deciding branching hyperproperties for real time systems}

\author{\IEEEauthorblockN{Nabarun Deka\IEEEauthorrefmark{1}, Minjian Zhang\IEEEauthorrefmark{1}, Rohit Chadha\IEEEauthorrefmark{2}, and Mahesh Viswanathan\IEEEauthorrefmark{1}}
\IEEEauthorblockA{\IEEEauthorrefmark{1}University of Illinois at Urbana-Champaign \\
\{ndeka2, minjian2, vmahesh\}@illinois.edu}
\IEEEauthorblockA{\IEEEauthorrefmark{2}University of Missouri, Columbia \\
chadhar@missouri.edu}
}

\begin{document}
\maketitle

\begin{abstract}
Security properties of real-time systems often involve reasoning about hyper-properties, as opposed to properties of single {\exec}s or trees of {\exec}s. These hyper-properties need to additionally be expressive enough to reason about real-time constraints. Examples of such properties include information flow, side channel attacks and service-level agreements. In this paper we study computational problems related to a branching-time, hyper-property extension of metric temporal logic (MTL) that we call {\hcmtl}. We consider both the interval-based and point-based semantics of this logic. The verification problem that we consider is to determine if a given {\hcmtl} formula $\hcmtlf$ is true in a system represented by a timed automaton. We show that this problem is undecidable. We then show that the verification problem is decidable if we consider {\exec}s upto a fixed time horizon $T$. Our decidability result relies on reducing the verification problem to the truth of an MSO formula over reals with a bounded time interval.
\end{abstract}


\section{Introduction}
Unlike the traditional safety and liveness properties, security guarantees such as non-interference desired of systems, are not \emph{trace-based} ~\cite{obs-mclean,hyperprop-clarkson}, and instead are properties of sets of executions. In their seminal paper, Clarkson and Schneider~\cite{hyperprop-clarkson}, called such requirements~\emph{hyperproperties}. Several temporal logics have been designed to express formally and reason about hyperproperties. By far, the most well-known of these logics are {\hypltl} and {\hypctl}~\cite{templogichyperprop-clarkson}. {\hypltl} and {\hypctl} extend the standard temporal logics {\ltl}~\cite{ltl-pnueli} and {\ctl}~\cite{ctl*-emerson} respectively. While {\hypltl} allows for reasoning about linear time hyperproperties, {\hypctl} allows for reasoning about branching time hyperproperties~\footnote{In linear time hyperproperties, the different executions being quantified are decided {\lq\lq}in advance{\rq\rq}, and in branching time hyperproperties, an execution being quantified may {\lq\lq}branch off{\rq\rq} in the middle of the last quantified execution. }. {\hypltl} and {\hypctl} differ from {\ltl} and {\ctl} by having explicit path variables and thus allowing for quantification over multiple executing traces simultaneously. The problems of checking a finite-state system against {\hypltl} and {\hypctl} formulas are shown to be decidable in~\cite{templogichyperprop-clarkson} by showing that the verification problem reduces to the satisfiability problem for quantified propositional temporal logic {\QPTL}~\cite{qptl-sistla}. {\QPTL} is a generalization of {\ltl}, and is interpreted over (untimed) transition systems. 

While {\hypltl} and {\hypctl} are able to express hyperproperties of transition systems, they are inadequate to express hyperproperties that relate {\lq\lq}timed{\rq\rq} executions, namely executions that are decorated by the \emph{time} at which each observation of the system occurs. Such hyperproperties, henceforth referred to as {\timhyp}, are essential to reason about the timing behaviors of a system. Some examples include the absence of timing leaks and timeliness of optimistic contract signing; see Section~\ref{sec:examples} for examples.
The need for reasoning about {\timhyp} has led to the development of timed hyperlogics, such as in~\cite{hyperstl,timedhyp,borzoo,khaza,gilles1,gilles2}. The real time system being analyzed in this context is usually modeled by \emph{timed} automata~\cite{alurdill} as in~\cite{hyperstl,timedhyp,khaza,borzoo}. In contrast, the logics in~\cite{gilles1,gilles2} are geared towards verifying timed properties of cryptographic properties, and models are timed versions of the applied-pi calculus~\cite{AbadiFournet2001}. (See Section~\ref{sec:related} on Pager~\pageref{sec:related} for a detailed discussion of  logics in ~\cite{gilles1,gilles2}.)

A timed automaton is a finite-state automaton augmented with a finite set of clocks. The clocks progress synchronously, and the automaton can make transitions based on their values and reset any of its clocks during a transition.  
Verifying timed systems is more challenging, even for regular, non-hyperproperties. Thus, in~\cite{borzoo,khaza}, the time model is taken to be discrete, \emph{i.e.}, the timed traces are sequences of pairs of the observed state of the system and the time observed, where the times are non-negative integers. 
In contrast,~\cite{timedhyp} considers verifying timed hyperproperties of timed automata when the time model is taken to be continuous/dense. In particular, they consider the \emph{point-based} semantics for timed automata~\cite{realtime-Alur,TPTL,Henzinger}. The timed traces in the point-based semantics are also sequences of pairs of observed states and observed times, but unlike ~\cite{borzoo,khaza}, the observations may occur at times that are arbitrary non-negative real numbers. For specifying {\timhyp},~\cite{timedhyp} extends the linear-time logic Metric Interval Temporal Logic ({\mitl})~\cite{mitl}   to Hyper Metric Interval Temporal Logic ({\hypmitl}) analogous to the extension of {\ltl} to {\hypltl}. {\mitl} is a commonly used logic to specify properties of timed systems, and is similar to {\ltl} except that the temporal modalities are annotated with non-singular\footnote{A non-empty interval is singular if it has exactly one element. Otherwise, it is non-singular. } time intervals: for example, $\phi\; U_I\; \psi$ means that $\psi$ must be true at some time $t\in I$ units from the current time and $\phi$ must hold at all times before $t.$   

The problem of verifying timed automata against {\hypmitl} specifications is undecidable~\cite{timedhyp}. The same paper also establishes that the verification problem becomes decidable when restricting to executions all of whose observations occur before a given time-bound $N.$ The decidability result is established by showing that the verification problem reduces to the satisfaction problem of {\QPTL}. Time-bounded verification is an often-adapted strategy for taming the complexity of verifying timed systems (See, for example,~\cite{boundedtime,worrelltalk}). Please note that even though time of the observations may be bounded, the total number of observations within that time-bound is not bounded.

\myparagraph{Our contributions} 
In this paper we consider the problem of verifying branching hyperproperties for timed automata. As has also been observed in~\cite{gilles1,gilles2}, several security guarantees for timed systems are branching hyperproperties; see Section~\ref{sec:examples}. In order to specify branching hyperproperties, we define the logic {\hcmtl} which is obtained by first extending the logic Metric Temporal Logic ({\mtl})~\cite{mtl}) to a branching-time logic (analogous to the extension of {\ltl} to {\ctl}) and then considering the {\lq\lq}hyper{\rq\rq} version of the resulting logic (analogous to the extension of {\ctl} to {\hypctl}). {\mtl} itself generalizes {\mitl} by allowing singular intervals to annotate the temporal operators. 
{The logic is presented in negation-normal form where negations are pushed down to propositions. Note that as a hyperproperty relates multiple executions, different timed traces in the hyperproperty may run for different times. This requires the logic to allow reasoning at times after the end of a trace. The negation normal form facilitates this when defining the semantics, as this allows for both a proposition and its negation to be false in a timed trace after it has ended. However, this means that we have to choose the temporal operators carefully to be able to express both the hyperproperty and its negation.}

In a departure from~\cite{timedhyp}, we also consider \emph{interval-based} semantics~\cite{mitl,interval-Henzinger,stateclock} for timed automata in addition to point-based semantics. In the interval-based semantics, the system modeled by a timed automaton is continuously under observation, and the timed traces of the automaton are a sequence of pairs of the observed state and the interval during which the state is observed. Verifying timed automata with interval-based semantics is more difficult than with point-based semantics. For example, the problem of verifying timed automata against {\mtl} specifications is undecidable for finite words in the interval-based semantics~\cite{Henzinger} but is decidable for finite words in the point-based semantics~\cite{WorrelMTL}.

We consider two verification problems for both interval-based and point-based semantics.
\vspace*{0.1cm}\\
\textbf{Bounded Time:} Given a timed automaton $\ta$, an {\hcmtl} specification $\hcmtlf$, and a time bound $N$, determine if $\hcmtlf$ is satisfied by $\ta$ when we consider only executions that are observed up-to time $N$.\\
\textbf{General:} Given a timed automaton $\ta$ and an {\hcmtl} specification $\hcmtlf$, determine if 
$\hcmtlf$ is satisfied by $\ta.$

\myparagraph{Bounded-Time verification}
Our first result is that the bounded-time verification is decidable for interval-based semantics. Note that as {\hypmitl} is a fragment of {\hcmtl}, this result generalizes the results of~\cite{timedhyp} to interval-based semantics. We make a few salient observations about the proof of this result. 

 Unlike~\cite{templogichyperprop-clarkson,timedhyp}, it is not clear that the verification problem can be reduced to {\QPTL} satisfaction for interval-based semantics. Instead, we choose to obtain decidability by showing the verification problem reduces to the problem of satisfaction of the Monadic Second Order Logic with strict inequality and successor ($\mso$), over the subset $[0,N)$ of reals. 
 {Specifically, we show that for each timed automaton $\ta$ and {\hcmtl} formula $\varphi$, there is a $\mso$ formula $\psi_{\ta,\varphi}$ with a set of free monadic predicates $\MP_{\ta}$  such that $\ta$ satisfies  $\varphi$ if and only if there is a  model $f$ that satisfies $\psi_{\ta,\varphi}$. A model of the formula $\psi_{\ta,\varphi}$, hereafter referred to as a \emph{flow}, is a function from the domain $[0,N)$ to the power-set of the set $\MP_{\ta}.$ Intuitively, for $t\in [0,N),$ $f(t)$ is  the set of predicates $P\in \MP_{\ta}$ that are true at $t.$ In order to establish this result, we carefully construct a one-to-one mapping from the {\exec}s of $\ta$ to the set of flows.}

 
Our next result is that the bounded-time verification is decidable for point-based semantics, thus generalizing the results of~\cite{timedhyp} to branching hyperproperties for point-based semantics.  The decidability result is established by showing that the bounded-time verification problem for point-based semantics is reducible to the bounded-time verification problem for interval-based semantics: for every timed automaton {$\ta_{\mathsf{pt}}$} and {\hcmtl} formula $\hcmtlf_{\mathsf{pt}}$, there is a timed automaton $\ta_{\mathsf{ib}}$ and a {\hcmtl} formula $\hcmtlf_{\mathsf{ib}}$ constructible from {$\ta_{\mathsf{pt}}$} and $\hcmtlf_{\mathsf{pt}}$ such that $\ta_{\mathsf{ib}}$ satisfies $\hcmtlf_{\mathsf{ib}}$ in the interval-based semantics if and only if  $\ta_{\mathsf{pt}}$ satisfies $\hcmtlf_{\mathsf{pt}}$ in the point-based semantics. The key observation is that both timed automata and {\hcmtl} in the interval-based semantics are at least as expressive as the point-based semantics.
In point-based semantics, we only consider time points where observations occur. Thus, we need to add a new proposition that marks these observations in the interval-based semantics. This reduction is valid even when the time domain is not bounded and for {\hypmitl} formulas.  

\myparagraph{General verification} 
Since verifying timed automata against {\mtl} semantics is already undecidable~\cite{Henzinger} for interval-based semantics, the general verification problem for {\hcmtl} is undecidable for interval-based semantics. Further, as the reduction of the verification problem for point-based semantics to the verification problem for interval-based semantics is also valid for {\hypmitl} formulas, and verifying timed automata against  {\hypmitl} specifications under point-based semantics is undecidable~\cite{timedhyp}, the undecidability carries over to both point-based and interval-based semantics even for {\hypmitl} specifications.\footnote{The undecidability proof in \cite{timedhyp} uses past operators. We do not have past operators in our logic. Nevertheless, we can also show that the problem is undecidable by reducing the universality problem of timed automata to {\hypmitl} verification under point-based semantics.}

\myparagraph{Organization} 
The rest of the paper is organized as follows. The logic {\hcmtl} and its interval-based semantics for timed automata is presented in Section~\ref{sec:logic}. Section~\ref{sec:examples} discusses examples of timed hyperproperties and their formalization in {\hcmtl}. Section~\ref{sec:decidability} presents the decidability proof of bounded-time verification in the interval-based semantics, and Section~\ref{sec:ptsemantics} presents the reduction from point-based semantics to interval-based semantics. Related work is discussed in Section~\ref{sec:related}, and we present our conclusions in Section~\ref{sec:conclusions}.

\ifdefined\arxiv
    This paper has been accepted to appear at the Computer Security Foundation 2024 conference.
\else
    Due to space constraints, we are not able to present the proofs in their complete technical details in this paper. Instead, we present the key ideas of the proofs and illustrate them using examples. The complete proofs can be found in the extended version of the paper on arxiv.
\fi
    
\section{The logic \hcmtl}
\label{sec:logic}

In this section, we introduce our logic {\hcmtl} which allows one to express branching hyperproperties of real time systems. {\hcmtl} is an extension of \emph{metric temporal logic} (\mtl)~\cite{mtl} which is a linear time logic to reason about real time constraints. Our real time systems will be modeled by timed automata~\cite{alurdill}, which is a popular model to describe such systems. We begin by defining the syntax of our logic, introduce timed automata, and conclude by formally defining what it means for a timed automaton to satisfy a formula in {\hcmtl}. As is standard, we use $\nat$ for natural numbers, $\reals$ for real numbers, $\nnreals$ for non-negative reals, and $\posreals$ for positive real numbers. 

\myparagraph{Intervals}
An \emph{interval} $I$ is of the form $\opb t_1, t_2 \cpb$ where $t_1, t_2 \in \reals \cup \{\infty\}$ with $t_1 \leq t_2$ ($\leq$ is defined as expected for $\infty$), and $\opb \in \{(,[\}$ and $\cpb \in \{),]\}$. We will denote by $t + \opb t_1,t_2 \cpb$, the interval $\opb t + t_1, t + t_2 \cpb$. For $I = \opb t_1,t_2 \cpb$, $t_1$ and $t_2$ are the \emph{left} (denoted $\lend{I}$) and \emph{right} (denoted $\rend{I}$) \emph{endpoints} of $I$, respectively. An interval is called \emph{singular} if it is of the form $[t,t]$. Two intervals $I_1 = \opb t_1, t_2 \cpb$ and $I_2 = \opb t_3, t_4 \cpb$ are said to be \emph{consecutive} if $t_2 = t_3$, $t_2$ is in exactly one of $I_1$ or $I_2$, and $I_1 \cap I_2 = \emptyset$. For example, $[1,2)$ and $[2,4)$ are consecutive intervals. An \emph{\is} is an finite sequence of intervals $I_1,I_2, I_3, \dots, I_n$ that satisfies the following two conditions: [Initial] $\lend{I_1} = 0$ and $0 \in I_1$; and [Consecution] for each $i \geq 1$, $I_i$ and $I_{i+1}$ are consecutive intervals.

\subsection{Syntax}
\label{sec:syntax}


Formulas in {\hcmtl} are built using atomic propositions $\props$, logic connectives, and modal operators. To allow one to reason about multiple {\exec}s, it has variables representing finite {\exec}s that can be quantified; $\pvar$ is the set of such \emph{path variables}. The BNF grammar for formulas in {\hcmtl} is given below.
\[
\hcmtlf \ ::= \ p_{\pi} \ | \ \neg p_{\pi} \ | \ \hcmtlf \vee \hcmtlf \ | \ \hcmtlf \wedge \hcmtlf \ | \ F_I \hcmtlf \ | \ G_I \hcmtlf \ | \ \hcmtlf U_{I} \hcmtlf \ | \  \exists \pi \hcmtlf \ | \ \forall \pi \hcmtlf
\]
In the grammar above, $p \in \props$ is an atomic proposition, $\pi \in \pvar$ is a path variable, and $I$ is an interval. As in MTL, we allow singular intervals of the form $[t,t]$ to decorate modal operators, which distinguishes it from other logics that are based on {\mitl}~\cite{mitl}. We use {\hypmitl} to denote the fragment of {\hcmtl} in which all quantifiers occur at the top-level of the formula and all intervals are non-singular.

Before defining the formal semantics for {\hcmtl}, we informally describe what formulas mean. The logic reasons about multiple finite {\exec}s of a real time system that are referred to by path variables. Atomic propositions capture abstract truths that may hold at different times during an {\exec}. In order to distinguish between propositions in different {\exec}s, we annotated propositions with the path for which it is being asserted. Thus, $p_\pi$ asserts that proposition $p$ is true in {\exec} $\pi$ currently. Similarly, $\neg p_\pi$ asserts that $p$ is false in $\pi$. It is important to recognize that here $\neg p_\pi$ asserts that $p$ is false, not that $p$ is not true. For example at a time $t$ that is after {\exec} $\pi$ ends, neither $p_\pi$ nor $\neg p_\pi$ hold for $p$ is neither true nor false as $\pi$ has ended. Thus, the law of excluded middle does not hold in {\hcmtl}. As a consequence, negation needs to be handled carefully (see Remark~\ref{rm:neg}) and consequently {\hcmtl} has more modal connectives than in typical presentations of a temporal logic.

Formulas can be combined using Boolean connectives conjunction ($\wedge$) and disjunction ($\vee$). The operators $F_I$, $G_I$, and $U_I$ express the modal and real-time aspects of the logic. These operators are real time extensions of the classical finally, globally, and until operators found in {\ltl}~\cite{ltl-pnueli} and {\ctl}~\cite{ctl*-emerson}, which constrain modal operators with time intervals requiring obligations to hold within those intervals. The formula $F_I \hcmtlf$ (read `finally $\hcmtlf$') asserts that $\hcmtlf$ holds at a time $t$ from the current time where $t$ is in the interval $I$. $G_I \hcmtlf$ (read `globally $\hcmtlf$') asserts that $\hcmtlf$ holds at all times $t$ from the current time when $t$ is in interval $I$. $\hcmtlf_1 U_I \hcmtlf_2$ (read `$\hcmtlf_1$ until $\hcmtlf_2$') holds if $\hcmtlf_2$ holds at some time $t$ units from the current time where $t \in I$ and $\hcmtlf_1$ holds continuously from the current time until $\hcmtlf_2$ holds. For example, $\hcmtlf_1 U_{[0,2]} \hcmtlf_2$ says that eventually, within $2$ units of time from now, $\hcmtlf_2$ is true and $\hcmtlf_1$ is true until then.

As mentioned before, {\hcmtl} expresses hyperproperties through path variables that stand for finite length {\exec}s, and quantifying over them. $\exists\pi \hcmtlf$ asserts that there is an {\exec} such that $\hcmtlf$ holds, while $\forall\pi \hcmtlf$ says that $\hcmtlf$ holds no matter what {\exec} is assigned to $\pi$. Like in other branching temporal logics for hyperproperties~\cite{templogichyperprop-clarkson}, when {\exec}s are quantified, the chosen {\exec} for the variable is required to be an extension or branch of a `current' {\exec}; the `current' execution depends on the context of the larger sentences in which a particular quantified sub-formula appears. This subtle aspect of quantification highlighted in Example~\ref{ex:quant}. The variable $\pi$ is \emph{bound} in the formulas $\exists\pi\hcmtlf$ and $\forall\pi\hcmtlf$. Any variable that does not appear within the scope of a quantifier is said to be \emph{free}. Without loss of generality, we assume that a variable is bound at most once and does not appear both bound and free; these assumptions can be easily met by renaming bound variables. Finally, due to technical reasons that will become clear when we discuss the formal semantics, the modal operators $F_I$, $G_I$, and $U_I$ are required to always appear within the scope of quantifier.

\begin{example}
    \label{ex:quant}
    As in other branching temporal logics for hyperproperties~\cite{templogichyperprop-clarkson}, quantified {\exec}s are required to be extensions or branches of certain other {\exec}s. We illustrate this through a couple of examples. The formula $\forall \pi_1 F_{[1,2]} \exists \pi_2 \hcmtlf$ says that for every {\exec} $\pi_1$ of the system, within $1$ to $2$ units of time, there is an extension $\pi_2$ of $\pi_1$ that satisfies $\hcmtlf$. In this case, the extension means that $\pi_1$ and $\pi_2$ must agree upto the point in time when the obligation for the operator $F$ is met. On the other hand, the formula $\forall \pi_1 \exists \pi_2 F_{[1,2]} \hcmtlf$ says that for every {\exec} $\pi_1$ of the system, there exists an {\exec} $\pi_2$ of the system such that within 1 to 2 units of time, $\pi_1$ and $\pi_2$ satisfy $\hcmtlf$. Here, $\pi_1$ and $\pi_2$ are not required to agree (except at the very beginning) and can be thought to be completely independent {\exec}s.
\end{example}

\begin{remark}
\label{rm:neg}
    It is typically convenient for logics to be closed under negation. Since negation in {\hcmtl} is restricted to only be applied to propositions, the logic has existential and universal quantifiers, and additional modal operators to ensure that negations can always be pushed inside. De Morgan's laws and the duality between exists and for all, allow one to handle the usual logical operators. For the modal connectives the following equivalences hold: $\neg F_I \hcmtlf \equiv G_I \neg \hcmtlf$, $\neg G_I \hcmtlf \equiv F_I \neg \hcmtlf$, and $\neg (\hcmtlf_1 U_I \hcmtlf_2) \equiv (G_I \neg \hcmtlf_2)\: \vee\: (\neg \hcmtlf_2 U_J \neg \hcmtlf_1)$ where $J = [0, R(I))$. 
\end{remark}
    
\subsection{Timed Automaton}
\label{sec:ta}

Timed automata~\cite{alurdill} are a popular formal model to describe real time systems. They are an extension of finite automata that are equipped with clocks that can be individually reset and measure time since the last reset. Clocks can be used to enforce real time constraints during a system execution. All clocks in a timed automaton are \emph{synchronous}, and progress in lockstep with an ambient global clock. We introduce this model in this section. 

\myparagraph{Clock Constraints}
A clock constraint over a clock $x$ is formula given by the following grammar.
\[
\cons \ ::= \ x \sim c \ | \ \cons \vee \cons \ | \  \cons \wedge \cons
\]
where $c \in \nat$ and $\sim \in \set{<, \leq, =, \geq, >}$. We will denote by $\conset{x}$ the set of all clock constraints over $x$. For a set of clocks $\clocks$, $\conset{\clocks}$ is the union of $\conset{x}$ for all $x \in \clocks$. A clock valuation is a map $\clockval: \clocks \rightarrow \nnreals$. The satisfaction relation $\clockval \models \cons$ is defined inductively as follows.
\begin{itemize}
    \item $\clockval \models x \sim c$ iff $\clockval(x) \sim c$ is true.
    \item $\clockval \models \cons_1 \vee \cons_2$ iff $\clockval \models \cons_1$ or $\clockval \models \cons_2$.
    \item $\clockval \models \cons_1 \wedge \cons_2$ iff $\clockval \models \cons_1$ and $\clockval \models \cons_2$.
\end{itemize}

\myparagraph{Timed Automata}
A \emph{timed automaton} over a set of atomic propositions $\props$ is a tuple, $\taext$, where:
\begin{itemize}
    \item $\st$ is a finite set of states.
    \item $\startst \subseteq \st$ is a set of initial states.
    \item $\stlab : \st \rightarrow \powerset{\props}$ is a state labeling function that labels each state with the set of propositions that are true at that state.
    \item $\clocks$ is a finite set of clocks.
    \item $\clockcons: \st \times \clocks \rightarrow \conset{\clocks}$ is a function that labels each state, clock pair $(v,x)$ with a clock constraint over $x$. 
    \item $\edges \subseteq \st \times \st \times \powerset{\conset{\clocks}}\times \powerset{\clocks}$ is a set of transition edges of the automaton. An edge $(v_1,v_2,\guardset,\gamma)$ is a transition from state $v_1$ to $v_2$ that satisfies all the clock constraints in the guard $\guardset$ and resets the clocks in $\gamma \subseteq \clocks$.
    \item $\final \subseteq \st$ is the set of final states.
\end{itemize}

\myparagraph{Runs/Executions}
An {\exec} $\run$ of $\ta$ is a finite sequence
\[
(v_1, I_1) \xrightarrow[\guardset_1]{\gamma_1}(v_2,I_2) \xrightarrow[\guardset_2]{\gamma_2}(v_3,I_3)\xrightarrow[\guardset_3]{\gamma_3}\dots \xrightarrow[\guardset_{n-1}]{\gamma_{n-1}}(v_n,I_n)
\]
where $v_i \in \st$ for all $i \in \{1,2,\dots,n\}$ and $I_1,I_2,\dots, I_n$ is an {\is} such that $v_1 \in \startst$, $(v_i, v_{i+1}, \guardset_i, \gamma_i) \in \edges$ for each $i$, and the real time constraints imposed in each state and transition are satisfied. To define when real time constraints are met, we need to define clock valuations at each time during the execution. We begin by first defining the clock valuations when first entering state $v_i$. Let the sequence of clock valuations $\clockval_1,\clockval_2, \dots \clockval_n$ be inductively defined as follows: $\clockval_1(x) = 0$ for all $x \in X$, and for all $i \geq 1$, $\clockval_{i+1}(x) = \clockval_{i}(x) + \rend{I_i} - \lend{I_i}$ if $x \not \in \gamma_i$, and $0$ otherwise. Next, for $t \in I_i$, the clock valuation at time $t$, $\clockval_t$, is given as $\clockval_t(x) = \clockval_i(x) + t - \lend{I_i}$. Finally, for $\run$ to be an execution the following two conditions must hold: [State Constraints] for every $i$, clock $x$ and $t \in I_i$, $\clockval_t(x) \models \clockcons(v_i, x)$, and [Guard Constraints] for every $i$, clock $x$, and $\guard \in \guardset_i \cap \conset{x}$, $\clockval'(x) \models \guard$, where $\clockval'(x) = \clockval_i(x) + R(I_i) - L(I_i)$.
\rmv{
    \begin{itemize}
        \item $v_1 \in \startst$. 
        \item For each $i \geq 1$, $(v_i, v_{i+1}, \guard_i, \gamma_i) \in \edges$.
        \item Define a sequence of clock valuations $\clockval_1,\clockval_2, \dots$ inductively as follows:
        \begin{itemize}
            \item $\clockval_1(x) = 0$ for all $x \in X$.
            \item For all $i \geq 1$, $\clockval_{i+1}(x) = \clockval_{i}(x) + \rend{I_i} - \lend{I_i}$ if $x \not \in \gamma_i$, and $\clockval_{i+1} = 0$ otherwise.
        \end{itemize}
        For $t \in I_i$, define a clock valuation $\clockval_t$ as $\clockval_t(x) = \clockval_i(x) + t - \lend{I_i}$. For each $i \geq 1$ and $t \in I_i$, $\clockval_t(x) \models \clockcons(v_i, x)$ must hold.
        \item For $i \geq 1$, define $t_i = R(I_i) = L(I_{i+1})$. If $t_i \in I_i$, then $\clockval_{t_i} \models \guard_i$ for all $\guard_i \in \guardset_i$. If $t_i \in I_{i+1}$, then the clock valuation $\clockval'$ defined as $\clockval'(x) = \clockval_i(x) + R(I_i) - L(I_i)$ must satisfy $\clockval' \models \guard_i$ for all $\guard_i \in \guardset_i$.
    \end{itemize}
}
    
The {\exec} $\run$ is said to be \emph{accepting} if $v_n \in \final$. The collection of all accepting {\exec}s of $\ta$ will be denoted by $\flows{\ta}$. We say $t \in \len{\run}$ iff $t \in I_i$ for some $i$. An accepting {\exec} $\run$ is said to be \emph{bounded by $N \in \reals$} if $t \not\in \len{\run}$ for all $t \geq N$; the set of all accepting {\exec}s bounded by $N$ of $\ta$ will be denoted by $\flowsn{\ta}$. For $t \in \len{\rho}$, the state at time $t$, denoted $\run(t)$, is $v_i$ if $t \in I_i$. The \emph{prefix of $\run$ up to time $t \in \len{\rho}$}, denoted $\run|_t$, is the {\exec}
\[
(v_1, I_1) \xrightarrow[\guardset_1]{\gamma_1}(v_2,I_2) \xrightarrow[\guardset_2]{\gamma_2}(v_3,I_3)\xrightarrow[\guardset_3]{\gamma_3}\dots \xrightarrow[\guardset_{i-1}]{\gamma_{i-1}}(v_i,J_t)
\]
where $t \in I_i$ and $J_t = I_i \cap [0,t]$. Two {\exec}s $\run_1$ and $\run_2$ are said to be \emph{equal up to time $t$} if $\run_1|_t$ and $\run_2|_t$ are identical.

\subsection{Semantics}
\label{sec:semantics}

We introduce the interval-based semantics here. Let us fix a timed automaton $\taext$. To define whether a {\hcmtl} formula $\hcmtlf$ is true in $\ta$ at a particular time $t$, we need to know what execution is assigned to each free path variable in $\hcmtlf$. This is captured by a \emph{path environment}. A path environment $\penv: \pvar \to \flows{\ta}$ is a mapping that associates with each path variable in $\pvar$ an accepting {\exec} of $\ta$; we assume that the set of free variables of $\hcmtlf$ is included in $\pvar$. When $\pvar = \emptyset$, the \emph{empty path environment} is denoted by $\emptypenv$. For a path environment $\penv$ over $\pvar$, a variable $\pi$, an {\exec} $\run \in \flows{\ta}$, $\penv[\pi \mapsto \run]$ denotes the path environment with domain $\pvar \cup \set{\pi}$ that is identical to $\penv$, except that $\pi$ is now mapped to $\run$. In branching hyperproperty logics like {\hypctl}~\cite{templogichyperprop-clarkson}, when a variable is quantified, it is expected to be assigned to an {\exec} that is an extension/branch of an {\exec} that is currently assigned to a variable (also see Example~\ref{ex:quant}). Therefore, to define the semantics, we also need to know which {\exec} needs to be extended next at a future quantification. This is captured by $\dagg$ which takes values in $\pvar \cup \set{\epsilon}$; when $\dagg = \epsilon$, it indicates that the next quantified variable is completely independent. The satisfaction relation for {\hcmtl} captures when a formula $\hcmtlf$ is true in a timed automaton $\ta$ at time $t$ with respect to a path environment $\penv$ and $\dagg \in \pvar \cup \set{\epsilon}$. It is denoted $\hcmtlmodels \hcmtlf$, and is defined inductively as follows.
\begin{itemize}
    \item $\hcmtlmodels p_\pi$ iff $t \in \len{\penv(\pi)}$, and $p \in \stlab(\penv(\pi)(t))$.
    \item $\hcmtlmodels \neg p_\pi$ iff $t \in \len{\penv(\pi)}$, and $p \not \in \stlab(\penv(\pi)(t))$.
    \item $\hcmtlmodels \hcmtlf_1 \vee \hcmtlf_2$ iff $\hcmtlmodels \hcmtlf_1$ or $\hcmtlmodels \hcmtlf_2$.
    \item $\hcmtlmodels \hcmtlf_1 \wedge \hcmtlf_2$ iff $\hcmtlmodels \hcmtlf_1$ and $\hcmtlmodels \hcmtlf_2$.
    \item $\hcmtlmodels F_I \hcmtlf$ iff there exists $t' > t$ such that $t' \in t + I$ and $\hcmtlargta{\penv}{t'}{\dagg}{\ta} \hcmtlf$.
    \item $\hcmtlmodels G_I \hcmtlf$ iff for all $t' > t$, such that $t' \in t + I$, $\hcmtlargta{\penv}{t'}{\dagg}{\ta} \hcmtlf$.
    \item $\hcmtlmodels \hcmtlf_1 U_I \hcmtlf_2$ iff there exists $t' > t$ such that $t' \in t + I$ and $\hcmtlmodelsarg{\penv}{t'}{\dagg} \hcmtlf_2$, and for all $t < t'' < t'$, $\hcmtlmodelsarg{\penv}{t''}{\dagg} \hcmtlf_1$. 
    \item $\hcmtlmodels \exists \pi \hcmtlf$ iff there is an {\exec} $\run \in \flows{\ta}$ such that $\hcmtlmodelsarg{\penv[\pi \mapsto \run]}{t}{\pi} \hcmtlf$ and either (a) $\dagg = \epsilon$ and $t = 0$, or (b) $t \in \len{\penv(\dagg)}$ and $\run|_t = \penv(\dagg)|_t$.
    \rmv{
        \begin{itemize}
            \item Either $\dagg = \epsilon$ and $t = 0$, or $\dagg \in \pvar$, $t \in \len{\penv(\dagg)}$, and  $\flow$ and $\penv(\dagg)$ are equal upto time $t$.
            \item Define a path environment $\penv'$ as $\penv'(\pi) = \flow$ and $\penv'(\pi') = \penv(\pi')$ for all $\pi' \neq \pi$.
            \item $\hcmtlmodelsarg{\penv'}{t}{\pi} \hcmtlf$. \mvc{Seems to be phrased awkwardly? I think $f$ should be pulled out of the sub-bullets since it is used in all.}
        \end{itemize}
    }
    \item $\hcmtlmodels \forall \pi \hcmtlf$ iff for every {\exec} $\run \in \flows{\ta}$, if either (a) $\dagg = \epsilon$ and $t = 0$, or (b) $t \in \len{\penv(\dagg)}$ and $\run|_t = \penv(\dagg)|_t$, then $\hcmtlmodelsarg{\penv[\pi \mapsto \run]}{t}{\pi} \hcmtlf$. 
    \rmv{
        \begin{itemize}
            \item Either $\dagg = \epsilon$ and $t = 0$, or $\dagg \in \pvar$ and $t \in \len{\penv(\dagg)}$, and if $\flow$ and $\penv(\dagg)$ are equal upto time $t$, the remaining points hold.
            \item Define a path environment $\penv'$ as $\penv'(\pi) = \flow$ and $\penv'(\pi') = \penv(\pi')$ for all $\pi' \neq \pi$.
            \item $\hcmtlmodelsarg{\penv'}{t}{\pi} \hcmtlf$. \mvc{Same comment as for exists.}\ndc{I do not like the phrasing for for all}
        \end{itemize}
    }
\end{itemize}

\rmv{
\begin{remark}
    Since this is a branching time logic, when we quantify over a path $\pi$, we look at runs $f$ that are exactly equal to $\dagg$ i.e. the last quantified path up to time  $t$. This exactly captures the idea that we quantify over all branches of the computation $\dagg$.
\end{remark}
}

\myparagraph{Bounded Time Semantics}
Bounded time semantics captures the notion that a timed automaton meets a specification $\hcmtlf$ up to time $N \in \nnreals$. This is captured by a satisfaction relation $\hcmtlmodelsbdd \hcmtlf$ which is defined in manner very similar to the definition above, except that $t$ is required to be $< N$, the path environment $\penv: \pvar \to \flowsn{\ta}$ maps variables to accepting {\exec}s bounded by $N$ and whenever a new variable is quantified, it is assigned an {\exec} in $\flowsn{\ta}$. The formal definition is skipped due to space constraints.

\myparagraph{Verification Problems}
This paper studies two decision problems associated with {\hcmtl}.\vspace*{0.1cm}\\
{\bf [General]} Given a timed automaton $\ta$ and an {\hcmtl} sentence $\hcmtlf$, determine if $\hcmtlmodelsarg{\emptypenv}{0}{\epsilon} \hcmtlf$.\\
{\bf [Bounded Time]} Given a timed automaton $\ta$, an {\hcmtl} sentence $\hcmtlf$, and a time bound $N$, determine if $\hcmtlmodelsargbdd{\emptypenv}{0}{\epsilon} \hcmtlf$.

\section{Examples}
\label{sec:examples}
In this section, we highlight the expressive power of {\hcmtl} through examples. Security specifications in these examples demand reasoning about multiple executions, have real-time constraints, require analyzing the branching structure, and use different quantifiers when bounding variables. The security guarantees in the first three examples (timing attacks, secure multi-execution, opacity) are linear hyperproperties. The security guarantees in the last four examples (timed commitments, contract signing, unlinkability, and fair reward) are branching hyperproperties that cannot be expressed in real-time extensions of {\hypltl}. 
Four examples involved quantifier alternation (opacity, contract signing, unlinkability, and fair reward). Two examples (timed commitment and contract signing) require non-trivial intervals to express the desired security guarantees. 
Finally, the unlinkability and fair reward examples are branching hyperproperties that relate executions along different branches and, thus, would not fall into branching time extensions of {\mtl}.
Thus the full power of {\hcmtl} is used to describe all the security requirements in these examples. Please note that in our examples below, we will use abbreviations like implication ($\implies$) and equivalence ($\iff$), which can be expressed in our logic by using the usual translation and pushing negations inside.

\myparagraph{Timing Attacks}
Programs computing over sensitive information should not be susceptible to leaking information through timing channels. To ensure that there are no such timing leaks, the program needs to guarantee that any two executions working on the same observable data (but possibly different private data) have the same timing behavior. Let $\cO$ be set of observable inputs, and let the proposition $\s{o}(a)$ for $a \in \cO$ denote that the input $a$ is observed. Let $\s{run}$ be the proposition to indicate that the program is running. Then using such propositions, the absence of timing leaks can be written as
\begin{align*}
    \forall \pi_1\forall \pi_2.( \bigwedge_{a \in \cO} \s{o}(a)_{\pi_1} & \iff \s{o}(a)_{\pi_2}) \implies\\
    & G_{[0,\infty)} (\s{run}_{\pi_1} \iff \s{run}_{\pi_2})
\end{align*}
This formula says that for all paths $\pi_1$ and $\pi_2$, if they start with the same observable inputs, then globally they should run for the same time.

\myparagraph{Secure Multi-Execution (SME)}
Non-interference requires that low-level (observable) outputs of two executions be the same if they are computing on the same low-level (observable) inputs. In other words, the difference in the high security inputs of two executions is not observable in the outputs. Secure multi-execution is an approach to ensure non-interference, where for any sequence of tasks, each task in the sequence is executed in two ways, one is a ``low copy'' and other is a ``high copy''. In the low copy, the high security inputs are set to some default values and the resulting outputs from this computation are observable. In the high copy, computation is carried with all the exact high security inputs, and the outputs from this computation are kept secure and non-observable. This ensures that any two executions operating on the same low-level inputs, have the same observations since only the outputs from the low copy are public which have default high security inputs. The low and high copies are interleaved for each task in the sequence. While SME ensures that the computation is non-interferent in the classical sense, it is open to timing attacks to an adversary observing the time duration between successive low copy computations if the high copy computations are of different length on inputs with the same low-level input. Such timing vulnerabilities can be described in {\hcmtl}. Let $\cO$ to be a set of observable input values and $\s{o}(a)$ for $a \in \cO$ be the proposition that low-level input $a$ is observed. Let $\s{Hstart}$, $\s{Hrun}$, $\s{Hend}$ be propositions denoting that high copy computation has started, is running, and has ended, respectively. Timing vulnerability can be written as
\[
\exists \pi_1\exists \pi_2.\ (\bigwedge_{a \in \cO} \s{o}(a)_{\pi_1} \iff \s{o}(a)_{\pi_2})\ U_{[0,\infty)}\ \hcmtlf
\]
where
\[
\begin{array}{rl}
    \hcmtlf = & \hspace*{-0.3cm} \s{Hstart}_{\pi_1} \wedge \s{Hstart}_{\pi_2} \wedge\\
        & \hspace*{-0.3cm}((\s{Hrun}_{\pi_1} \wedge \s{Hrun}_{\pi_2})\ U_{[0,\infty)}\ (\neg (\s{Hend}_{\pi_1} \iff \s{Hend}_{\pi_2})))
\end{array}
\]
The formula says that there {\exec}s $\pi_1$ and $\pi_2$ that have the same low-level inputs until time $t$ when the formula $\hcmtlf$ becomes true. $\hcmtlf$ asserts that at time $t$ high copy computations start in $\pi_1$ and $\pi_2$ and the computation in either $\pi_1$ or $\pi_2$ ends before the other.

\myparagraph{Opacity}
Opacity of a property $\psi$ demands that the truth of $\psi$ be undeterminable to an adversary. This can be formalized by demanding that for every {\exec} $\pi_1$ there is another {\exec} $\pi_2$ that has the same observable behavior, but $\psi$ is true in exactly one of $\pi_1$ and $\pi_2$. To state this in {\hcmtl}, we use the following propositions: for a set of observations $\cO$, the proposition $\s{o}(a)$ asserts that $a \in \cO$ is observed; $\s{end}$ asserts that the computation has ended; and, $\psi$ is property of the last state expressed using a Boolean combination of some propositions. Then opacity of $\psi$ is
\[
\begin{array}{rl}
    \forall \pi_1 \exists \pi_2.\ G_{[0,\infty)} & \hspace*{-0.3cm}((\bigwedge_{a \in \cO} \s{o}(a)_{\pi_1} \iff \s{o}(a)_{\pi_2}) \\ 
    & \hspace*{-0.5cm}\wedge ((\s{end}_{\pi_1} \wedge \s{end}_{\pi_2}) \implies \neg (\psi_{\pi_1} \iff \psi_{\pi_2})))
\end{array}
\]
Notice that opacity requires alternation of path quantifiers in order to express it.

\myparagraph{Timed Commitment}
Consider the problem of tossing a coin when the \emph{caller} (Alice) and the \emph{tosser} (Bob) are in different locations but have a \emph{reliable} communication channel to use. The setup is that Alice calls the toss, Bob tosses, and Alice wins if the result of the toss is what she predicts while Bob wins if it is not. A n\"{a}ive  protocol might be that Alice sends her prediction to Bob and Bob then sends the result of the coin toss to Alice. At this point both parties know who the winner is. However, such a protocol is not fair to Alice as a dishonest Bob can always report a result of a coin toss that is the opposite of what Alice called. To circumvent this, Alice could \emph{commit} her call, instead of sending it. Bob then tosses his coin and shares the result with Alice. At this point, Alice reveals her commitment and now both parties know the winner. However, such a protocol could be unfair to Bob, as a dishonest Alice may never reveal her commitment if she realizes that she lost. The solution that ensures fairness for both parties is to use a \emph{timed commitment}, where the commitment is revealed to Bob after the elapse a fixed time $T$ even if Alice takes no steps towards revealing her commitment. The steps of such a protocol are as follows: Alice commits her call within time $t_c$, Bob has to share the result of the coin toss within time $T$ after Alice's commitment, and if he does, either Alice will reveal her call or Bob can compute what Alice committed to after $T$ units after Alice's commitment. If Bob does not toss a coin within time $T$, Alice is released from her commitment. Let us formally define the fairness guarantees for Alice and Bob, which are different since the protocol is asymmetric. Let the proposition $\s{c}(b)$, for $b \in \set{\s{H},\s{T}}$ denote that Alice has committed bit $b$, even though $b$ itself is not revealed to Bob. We assume that communication is reliable and so once Alice commits, Bob knows that she did. Proposition $\s{t}(b)$ for $b \in \set{\s{H},\s{T}}$ denotes a state when Bob has shared the result of a coin toss to be $b$. Finally $\s{r}(b)$ asserts that Alice's commitment has been revealed to Bob and that call was $b$. Fairness for Bob now is
\[
\forall \pi_1.\ G_{[0,t_c]} \bigwedge_{b \in \set{\s{H},\s{T}}} (\s{c}(b)_{\pi_1} \implies (\forall \pi_2.\ F_{[T,\infty)} \s{r}(b)_{\pi_2})).
\]
It says for every {\exec} $\pi_1$, if Alice commits within time $t_c$, then in every extension $\pi_2$ of $\pi_1$, Bob will be able to see Alice's commitment at some time that is $T$ units after Alice's commitment. In addition, Bob will see the same bit that Alice committed. Fairness for Alice can be written as
\[
\begin{array}{rll}
\forall \pi_1.\ G_{[0,t_c]} & ((\s{c}(\s{H})_{\pi_1} \vee \s{c}(\s{T})_{\pi_1}) \implies \\
 & \forall \pi_2.\ G_{[0,T]} ((\neg \s{t}(\s{H})_{\pi_2} \wedge \neg \s{t}(\s{T})_{\pi_2}) \implies \\
 & \hspace*{0.71in}(\neg \s{r}(\s{H})_{\pi_2} \wedge \neg \s{r}(\s{T})_{\pi_2})))
\end{array}
\]
which says that in every {\exec}, if Alice commits within time $t_c$ then for the next $T$ units, if Bob has not yet shared the result of his coin toss, then Alice's commitment is not revealed. Notice that these properties are branching hyperproperties that cannot be expressed in linear hyperproperty logics even if they are real time.

\myparagraph{Contract Signing}
Consider a  contract signing protocol mediated by a trusted third party (TTP) where two parties wish to sign a contract. Each signer must have the ability to move on in a timely manner if the other party does not complete the signing protocol. Each signer must be able to reach an abort state if they want or must receive a message from the TTP that the protocol is aborted in a timely manner. For $i \in \set{1,2}$, let us consider the following propositions. $\s{start}(i)$ means that party $i$ has started the protocol, $\s{signed}(i)$ means that party $i$ has a signed contract, $\s{abort}(i)$ means that party $i$ is in an abort state, and $\s{token}(i)$ means that the TTP has provided an abort token to party $i$. We can now write the fairness for party $i$ as
\[
\begin{array}{rl}
\forall \pi_1.\ G_{[0,t]} & \hspace*{-0.3cm}(\s{start}(i)_{\pi_1} \implies\\
 & \hspace*{-0.3cm}(\exists \pi_2.\ F_{[0,T]} (\s{signed}(i)_{\pi_2} \vee \s{abort}(i)_{\pi_2}) \\
 &\hspace*{1.7cm}\vee \,\s{token}(i)_{\pi_2}))).
\end{array}
\]
This is once again a branching hyperproperty and it also involves quantifier alternation.

\myparagraph{Unlinkability} Radio Frequency Identification (RFID) is a technology used to identify and track physical objects using electromagnetic tags. A RFID system consists of tags attached to objects of interest and a reader that can communicate with the tags using electromagnetic waves to ascertain their identity and location. A security property that is required in such systems is unlinkability \cite{unlinkability} i.e. when multiple rounds of communication happens between a tag and the reader, an adversary that is observing the communications should not be able to link the communications to the same tag. If this property is violated, an adversary can potentially track an object by linking all communications between that particular tag and the receiver. Consider a simple RFID system with two tags. For $i \in \set{1,2}$, let $\s{comm}(i)$ denote that tag $i$ communicated with the receiver. For a set $\cO$ of observations, let the propositions $\s{o}(a)$ denote that $a \in \cO$ has been observed. We can state unlinkability of communications between tag $1$ and the receiver in \hcmtl{} as
\[
\begin{array}{ll}
     \forall \pi_1.(\s{comm}(1)_{\pi_1} \implies & \\
     \hspace*{0.78cm}G_{[0,\infty)} (\forall \pi_2 \exists \pi_3. &\hspace*{-0.5cm}F_{(0,\epsilon)}(\s{comm}(1)_{\pi_2}) \implies \\ &F_{(0,\epsilon)}(\s{comm}(2)_{\pi_3}) \wedge \varphi)).
\end{array}
\]
where $\epsilon > 0$ and
\[
\varphi = G_{[0,\infty)}(\bigwedge_{a \in \cO} \s{o}(a)_{\pi_2} \iff \s{o}(a)_{\pi_3}).
\]
It says that for any execution $\pi_1$ where tag 1 communicates with the receiver initially, globally, for all branches $\pi_2$ where tag 1 communicates with the receiver, there is another branch $\pi_3$ where tag 2 communicates with the receiver, such that $\pi_2$ and $\pi_3$ have the same observable behaviour to an adversary. This ensures that the adversary cannot link the two communications between tag 1 and the receiver. Observe that this is a branching \emph{hyperproperty} that relates two traces, and hence would not be captured by a branching time extension of MTL.

\myparagraph{Fair Reward}
Consider a distributed program running on top of a blockchain based cryptocurrency as described in \cite{gilles2}. The system consists of users that submit transactions in the form of contracts to the program, which are then executed and published by the program to the blockchain. In between a transactions submission and publication, it becomes public to other users, who might then choose to submit new transactions based on this. Note that we are considering a distributed program running on top of the blockchain and hence the submissions are not necessarily published in order of submission. The only requirement of the program is that submissions are eventually published or returned as invalid contracts. Consider a simple such model where a user if rewarded (monetarily) for correctly computing the pre-image of some hashed value. In such a system, after an honest user submits a correct answer (in the form of a transaction), an adversary might use the information from the transaction to submit their own answer and if it gets published first, the adversary can steal a promised reward from the honest user. The fair reward property requires that the program should not be vulnerable to such attacks. Formally, for any execution of the system, there is another execution (called ideal execution) where submissions are published in order and the final balances of all users in both executions are equal. The protocol allows a setup phase for the adversary where the adversary chooses the attack parameters. The proposition $\s{setup}$ indicates the end of the setup phase. The fair reward property now requires that in any execution of the system, once the adversary is finished setting up, there is an ideal branch of the execution such that both executions have the same balance for all users eventually. For a set $\cT$ of transactions (each transaction can be thought of as a bit string) and a transaction $x \in \cT$, let $\s{submit}(x)$ and $\s{publish}(x)$ denote that $x$ has been submitted and published respectively. Let $\s{silent}$ denote that no transaction has been submitted or published. In the remainder of this example, to avoid clutter, we will use the symbols $F,G,U$ without any interval annotation to indicate that the interval is $[0,\infty)$. An ideal execution can be modeled in \hcmtl{} as
\[
\begin{array}{rl}
    \s{ideal}(\pi) = & \hspace*{-0.25cm}G\, (\bigwedge_{x \in \cT} \s{submit}(x)_{\pi} \wedge F \,\s{publish}(x)_\pi \\
     & \hspace*{1cm} \implies \s{silent}_\pi U \s{publish}(x)_\pi). 
\end{array}
\]
This says that globally, if a transaction is submitted and eventually published, then the execution is silent until the publication. Let $\cB$ be the set of all possible balance profiles (these can be thought of as bit strings indicating the balance of each user), and for $b \in \cB$, let $\s{balance}(b)$ be the proposition indicating that the balance profile is described by $b$.
Fair reward can now be modeled  in \hcmtl{} as
\[
\begin{array}{rl}
     \forall \pi_1.\, G(&\hspace*{-0.3cm}\s{setup}_{\pi_1} \implies \\
     &\exists \pi_2.\, \s{ideal}(\pi_2) \wedge  \\
     & (\bigwedge_{x \in \cT} F\, \s{submit}(x)_{\pi_1} \iff F\, \s{submit}(x)_{\pi_2}) \wedge\\
     &F \, (G\,  (\s{silent}_{\pi_1} \wedge \s{silent}_{\pi_2}) \wedge \\
     & \hspace*{0.3cm}\bigwedge_{b \in \cB} \s{balance}(b)_{\pi_1} \iff \s{balance}(b)_{\pi_2}))     
\end{array}  
\]
This says that for every execution $\pi_1$, globally, once the adversary is done setting up, there is a branch $\pi_2$ which is an ideal execution, and the submissions in $\pi_1$ and $\pi_2$ are identical, and eventually, when both executions become silent, they have the same balance profiles.
\section{Verifying \hcmtl}
\label{sec:decidability}

In this section we present the main results related to the verification problems for {\hcmtl} introduced in Section~\ref{sec:semantics}. We show that the general verification is undecidable, but the bounded time verification problem is decidable. All the results we present in this section are for interval-based semantics. 

\begin{theorem}
    \label{thm:undec}
    The general verification problem for {\hcmtl} is undecidable in the interval-based semantics. In fact the verification problem is undecidable even for the fragment \hypmitl{}.
\end{theorem}
The proof of Theorem~\ref{thm:undec} is deferred to Section \ref{sec:ptsemantics}.

\begin{theorem}
    \label{thm:bdd-dec}
    The bounded time verification problem for {\hcmtl} is decidable in the interval-based semantics.  
\end{theorem}

Our decidabiliy result is established by reducing the bounded time verification problem for {\hcmtl} to the satisfiability problem for Monadic Second Order logic with $<$ and $+1$ relations, denoted $\mso$, over a bounded time domain. The satisfiability problem for $\mso$ is decidable over bounded time domains~\cite{boundedtime}, and our result thus follows. The following subsections outline the decidability proof and it is organized as follows.
\begin{enumerate}
    \item First we introduce $\mso$ and state the relevant decidability results.
    \item The first technical result in our proof shows that for any timed automaton $\ta$, there is an $\mso$ formula $\msof_\ta$ whose models are exactly the accepting {\exec}s of $\ta$.
    \item Finally, using this translation of a timed automaton to an $\mso$ formula, we reduce the bounded time verification problem to the satisfiability problem of $\mso$.
\end{enumerate}

\subsection{Monadic Second Order Logic}
\label{sec:mso}

Monadic Second Order logic with $<$ and $+1$, denoted $\mso$, is built over a set of monadic predicates {\MP} and set of first order variables $\freevar$. The BNF grammar is as follows. 
\[
\msof \ ::= \ x < y \ | \ +1(x,y)\ | \ P(x) \ | \ \msof \vee \msof \ | \ \neg \msof \ | \ \exists x \msof \ | \ \exists P \msof
\]
In the grammar above, $P \in \MP$ is a monadic predicate, and $x \in \freevar$ is a first order variable.

The semantics of $\mso$ is defined over a timed domain $\tdomain$. We will define it over two time domains $\tdomain = \nnreals$ and $\tdomain = [0,N)$ for some fixed $N \in \nat$. Let $\MPP \subseteq \MP$ be the set of free monadic predicates in an $\mso$ formula $\msof$. A \emph{flow} is a map, $\flow: \tdomain \rightarrow \powerset{\MPP}$ that is \emph{finitely variable} (explained next). For any interval $I \subseteq \tdomain$, define the map $\flow|_I: I \rightarrow \powerset{\MPP}$ as $\flow|_I(t) = f(t)$ i.e. it is the restriction of $\flow$ to the interval $I$. Now, we say a map $\flow : \tdomain \rightarrow \powerset{\MPP}$ is \emph{finitely variable} if for any bounded interval $I \subseteq \tdomain$ with finite endpoints, $\flow|_I$ has finitely many discontinuities i.e. the values of the monadic predicates change finitely many times in $I$. For $\textsf{Q} \subseteq \MPP \subseteq \MP$, and a flow $\flowg:\tdomain \rightarrow \powerset{P}$, we will denote by $g|_{\textsf{Q}}: \tdomain \rightarrow \powerset{Q}$ the flow defined as $g|_{\textsf{Q}}(t) = g(t) \cap \textsf{Q}$. An interpretation $\inter$ is a map $\inter: \freevar \rightarrow \tdomain$. We will denote by $\inter[x \mapsto a]$ the interpretation that maps $x$ to $a \in \tdomain$, and is same as $\inter$ for all $y \neq x$. The semantics, denoted by $\flow, \inter \models \msof$, is defined as follows.
\begin{enumerate}
    \item $\flow, \inter \models x < y$ iff $\inter(x) < \inter(y)$.
    \item $\flow, \inter \models +1(x,y)$ iff $\inter(y) = \inter(x) + 1$.
    \item $\flow, \inter \models P(x)$ iff $P \in \flow(\inter(x))$.
    \item $\flow, \inter \models \msof_1 \vee \msof_2$ iff $\flow, \inter \models \msof_1$ or $\flow, \inter \models \msof_2$.
    \item $\flow, \inter \models \neg \msof$ iff $\flow, \inter \not \models \msof$.
    \item $\flow, \inter \models \exists x\msof$ iff there exists $a \in \tdomain$ such that $\flow, \inter[x \rightarrow a] \models \msof$.
    \item $\flow, \inter \models \exists Q \msof$ iff there is some finitely variable flow $\flowg: \tdomain \rightarrow \powerset{\MPP \cup \set{Q}}$ such that $\flowg|_\MPP = \flow$ and $\flowg, \inter \models \msof$.
\end{enumerate}

\myparagraph{Satisfiability Problem}
Given an $\mso$ formula $\msof$ over a set $\MP$ of free monadic predicates and free first order variables $\freevar$, determine if there is a flow $\flow: \tdomain \rightarrow \powerset{\MP}$ and $\inter: \freevar \rightarrow \tdomain$ such that $\flow, \inter \models \msof$.

When the time domain is bounded, i.e., $\tdomain = [0,N)$ for some $N \in \posreals$, the satisfiability problem is decidable~\cite{boundedtime}.
\begin{theorem}[\cite{boundedtime}]
    \label{thm:mso-dec}
    For $\tdomain = [0,N)$ ($N \in \posreals$), the satisfiability problem for $\mso$ is decidable. 
\end{theorem}

\subsection{Translating Timed Automata to $\mso$}
\label{sec:AutomataToMSO}

We will now show that for any timed automaton $\ta$, there is an $\mso$ formula $\msof_\ta$ that is satisfied exactly by the accepting {\exec}s of $\ta$. The first challenge in this translation is that $\mso$ models are flows, which are functions from $\tdomain$ to a set of monadic predicates, which are different from {\exec}s of a timed automaton. Hence, to formally state our result, we need to construct a one-to-one correspondence between {\exec}s of a timed automaton and flows.

Recall that an {\exec} $\run$ of a timed automaton $\taext$ is of the form
$$(v_1, I_1) \xrightarrow[\guardset_1]{\gamma_1}(v_2,I_2) \xrightarrow[\guardset_2]{\gamma_2}(v_3,I_3)\xrightarrow[\guardset_3]{\gamma_3}\dots \xrightarrow[\guardset_{n-1}]{\gamma_{n-1}}(v_n,I_n)$$
We will encode this as a flow $\flow : \nnreals \rightarrow \powerset{V}$ over the set of monadic predicates $\MP = \st$ as follows:
\begin{itemize}
    \item If $t \in I_i$, $\flow(t) = \{v_i\}$
    \item If $t \not \in I_i$ for any $i$, $\flow(t) = \emptyset$ which indicates that the run has terminated and hence no state is present in the flow.
\end{itemize}
This is not sufficient because this does not give a one-to-one mapping from {\exec}s to flows. As an example, consider the following two simple {\exec}s:
\begin{align*}
    &\run_1 = (v, [0,10]) \\
    &\run_2 = (v, [0,5])\xrightarrow[\guardset]{\gamma}(v, (5,10]).
\end{align*}
These {\exec}s are different because one has a transition at time $5$ while the other does not. However, if we think of them as a function $\flowrun: [0,10] \rightarrow \st$, they are identical. This indicates that in the function $\flow$, we also need to carry information about the transitions and the clock resets that occur during an {\exec}. This involves some challenging subtleties that we illustrate with an example. Consider the following {\exec} with two clocks $x_1$ and $x_2$,
\begin{align*}
    \run = (v_1,[0,5)) &\xrightarrow[\guardset_1]{\{x_1\}} (v_2, [5,10]) \xrightarrow[\guardset_2]{\{x_2\}} (v_3, (10,12)) \\
    &\xrightarrow[\guardset_3]{\{x_2\}} (v_1, [12,12]) \xrightarrow[\guardset_4]{\{x_1\}} (v_4, (12,15))
\end{align*}
Observe that at time $5$, a transition $e_1 = (v_1,v_2,\guardset_1,\{x_1\})$ occurs and the automaton is in the target state of the transition i.e. state $v_2$. Let us call this type of transition a $\Tez$ transition. At time $10$ on the other hand, a transition $e_2 = (v_2,v_3,\guardset_2,\{x_2\})$ occurs and the automaton is in the source state of the transition i.e. state $v_2$, which we will call a $\Teo$ transition. Finally, at time $12$, both kinds of transitions occur. Thus, in the function $\flowrun$, we will define $\flowrun(5) = \{v_2, \Tez_{e_1}\}$ to say that the {\exec} is at state $v_2$ and a transition $e_1$ of type $\Tez$ occurred. Similarly we can define $\flowrun(10) = \{v_2, \Teo_{e_2}\}$ and at time $12$, and $\flowrun(12)$ will contain the location $v_1$ and one $\Tez$ and $\Teo$ transitions. Corresponding to each transition, some clocks are reset. Now in a singular interval, for example time $12$ in the {\exec} above, $x_2$ is reset at the transition $(v_3,v_1, \guardset_3, \{x_2\})$ and then $x_1$ is reset in the transition $(v_1,v_4,\guardset_4,\{x_1\})$. Hence, we also need to differentiate clock resets into two types, resets associated to $\Tez$ transitions and those associated to $\Teo$ transitions. So for each clock $x \in X$, we will have two monadic predicates $\rz{x}$ and $\ro{x}$. When a $\Teo$ transition corresponding to edge $e = (v_1, v_2, \guardset, \gamma)$ occurs at time $t$, we will add the resets $\rz{x}$ for $x \in \gamma$ to $\flowrun(t)$, and similarly for $\Teo$ transitions. The corresponding flow for the {\exec} above will be:
    \begin{itemize}
        \item $\flowrun(t) = \{v_1\}$ for $t \in [0,5)$
        \item $\flowrun(5) = \{v_2, \Tez_{e_1}, \rz{x_1}\}$
        \item $\flowrun(t) = \{v_2\}$ for $t \in (5,10)$
        \item $\flowrun(10) = \{v_2, \Teo_{e_2}, \ro{x_2}\}$
        \item $\flowrun(t) = \{v_3\}$ for $t \in (10,12)$
        \item $\flowrun(12) = \{v_1, \Tez_{e_3}, \rz{x_2}, \Teo_{e_4}, \ro{x_1}\}$
        \item $\flowrun(t) = \{v_4\}$ for $t \in (12,15)$
    \end{itemize}
This information is sufficient to ensure a one-to-one correspondence between {\exec}s and their corresponding flows. Using this idea, formally, we have the lemma
\begin{lemma}
    \label{runtoflow}
    For a timed automaton $\taext$, let $T = \setpred{\Tez_e, \Teo_e}{e \in \edges}$ and $R = \setpred{\rz{x}, \ro{x}}{x \in \clocks}$. Define a set of monadic predicates $\MP = \st \cup T \cup R$. Define $\cal{F}$ to be the set of all flows $f:  \nnreals \rightarrow \powerset{\MP}$. There is a one-to-one encoding $\runtoflow{\ta} : \flows{\ta} \rightarrow \cal{F}$ of {\exec}s $\run$ of $\ta$ as flows.
\end{lemma}
\ifdefined\arxiv
    \begin{proof}
        An {\exec} $\run$ of a timed automaton $\taext$ is of the form
$$(v_1, I_1) \xrightarrow[\guardset_1]{\gamma_1}(v_2,I_2) \xrightarrow[\guardset_2]{\gamma_2}(v_3,I_3)\xrightarrow[\guardset_3]{\gamma_3}\dots \xrightarrow[\guardset_{n-1}]{\gamma_{n-1}}(v_n,I_n)$$

    We define a flow $f: \nnreals \rightarrow \powerset{\MP}$ as follows:
    \begin{itemize}
        \item For $t \not \in I_i$ for any $i \in \{1,2,\dots,n\}$, define $f(t) = \emptyset$.
        \item For $t \in I_i$ with $t \neq R(I)$ and $t \neq L(I)$, define $f(t) = \{v_i\}$.
        \item For a time $t$ at which one transition occurs, if it is a $\Tez$ transition $e = (v, v', \guardset, \gamma)$, define $f(t) = \{v'\} \cup \{\Tez_e\} \cup_{x \in \gamma} \{\rz{x}\}$.
        \item For a time $t$ at which one transition occurs, if it is a $\Teo$ transition $e = (v, v', \guardset, \gamma)$, define $f(t) = \{v\} \cup \{\Teo_e\} \cup_{x \in \gamma} \{\ro{x}\}$.
        \item For a time $t$ at which two transitions occur, one is of type $\Tez$, lets call it $e_1 = (v,v',\guardset_1,\gamma_1)$, and the other is of type $\Teo$, lets call it $e_2 = (v',v'',\guardset_2,\gamma_2)$. Define $f(t) = \{v'\} \cup \{\Tez_{e_1}, \Teo_{e_2}\} \cup_{x \in \gamma_1} \{\rz{x}\} \cup_{x\in \gamma_2} \{\ro{x}\}$.
    \end{itemize}
    Define the map $\runtoflow{\ta}$ as $\runtoflow{\ta}(\run) = f$. Pick any two {\exec}s $\run_1$ and $\run_2$ of $\ta$ such that $\run_1 \neq \run_2$. Then, there is a time point $t$ where either the {\exec}s are at different states of the automaton $\ta$, or the {\exec}s take different transitions. In both cases, $\runtoflow{\ta}(\run_1)(t) \neq \runtoflow{\ta}(\run_2)(t)$, and thus, $\runtoflow{\ta}(\run_1) \neq \runtoflow{\ta}(\run_2)$.
    \end{proof}
\else

\fi
If we consider bounded {\exec}s of $\ta$ for some time bound $N \in \posreals$, the same encoding gives an encoding of bounded {\exec}s as flows $\flow: [0,N) \rightarrow \powerset{\MP}$.

    Now, given a timed automaton $\ta$, we want to construct an $\mso$ formula $\msof_\ta$ over the monadic predicates $\MP = \st \cup T \cup R$ such that $\msof_\ta$ is satisfied exactly by the set $\runtoflow{\ta}(\flows{\ta})$. The key idea behind this is that all the properties of an {\exec} can be expressed in $\mso$. We illustrate how certain important properties of an {\exec} can be expressed in $\mso$.

    The most basic property of an {\exec} is that it terminates at some time $l$ and at any time $t$ up to time $l$, the {\exec} is exactly in one state of the automaton. First we have a formula $US(t)$ ($US$ stands for Unique state) which states that there is exactly one state at time $t$.
    \begin{align*}
        US(t) = \bigvee_{v \in \st} v(t) \wedge \bigwedge_{v_1,v_2 \in \st: v_1 \neq v_2}\neg (v_1(t) \wedge v_2(t))
    \end{align*}
    Now, using this, we can express the above property as a disjunction of the following two formulas
    \begin{align*}
        \exists l\forall t((t \leq l \implies US(t)) \wedge (t > l \implies \wedge_{v \in \st} \neg v(t))) \\
        \exists l\forall t((t < l \implies US(t)) \wedge (t \geq l \implies \wedge_{v \in \st} \neg v(t)))
    \end{align*}
    The first one says that for time $t$ up to and including time $l$, there is exactly one state and after $l$ there is no state. The  second one says that for time $t$ up to but not including $l$, there is exactly one state and for time $t \geq l$, there is no state. In a similar manner we also say that at all times where the {\exec} is in some state,  at most one $\Tez_e$ and one $\Teo_e$ predicate is true.

    Another important property of an {\exec} is that if a transition $e \in \edges$ of type $\Tez$ occurs at time $t$, then the predicate $\Tez_e$ should be true at time $t$ and vice versa. For each $e = (v_1,v_2,\guardset, \gamma)$ we have the formula
        \begin{align*}
            \forall t&( \Tez_e(t)\implies v_2(t) \\
            &\wedge \exists y(y < t \wedge \forall z (y < z < t \implies v_1(t)))).
        \end{align*}
    This ensures that if $\Tez_e$ is true at time $t$, then there is transition of the form 
    $$(v_1, \opb t_1, t)) \xrightarrow[\guardset]{\gamma} (v_2, [t, t_2\cpb)$$ at $t$. For the opposite direction, i.e. to ensure that $\Tez_e$ is true only at times $t$ where a corresponding transition $e$ occurs, for each pair of states $v_1 \neq v_2$, let $C$ be the set of all edges of the form $e = (v_1, v_2, \guardset, \gamma)$. We have the formula
    \begin{align*}
        \forall t(&v_2(t)\wedge \exists y(y < t \wedge \forall z (y < z < t \implies v_1(t))) \\
        &\implies \bigvee_{e \in C} \Tez_e(t) \wedge \bigwedge_{e_1,e_2 \in C: e_1 \neq e_2} \neg (\Tez_{e_1} \wedge \Tez_{e_2}))
    \end{align*}
    This formula says that if for a small open interval $(y,t)$ the automaton is in state $v_1$ and at time $t$ it enters state $v_2$, then exactly one transition of the form $(v_1,v_2, \guardset, \gamma)$ must have occurred at time $t$. An important point to note here is that the second formula is only for edges where $v_1 \neq v_2$. This is because if $v_1 = v_2$, i.e. the edge is a self loop, then it is not necessary for a transition to occur. The automaton may simply remain in state $v_1$ without making a transition at time $t$. We can write similar formulas for transition of type $\Teo$.

    We also need to ensure that clock resets happen only when transitions occur. For this, first we ensure that if no transition occurs at time $t$, then no clocks are reset at that time.
    \begin{align*}
        \forall t (\neg (\vee_{e \in \edges} \Tez_e(t)) \implies \neg (\vee_{x \in \clocks} \rz{x}(t)))
    \end{align*}
    For the other direction, we say that if a transition $e = (v_1,v_2, \guardset, \gamma)$ of type $\Tez$ occurs, then the corresponding clock resets of type $\rz{x}$ occur.
    \begin{align*}
        \forall t(\Tez_e(t) \implies \wedge_{x \in \gamma} \rz{x}(t) \bigwedge \wedge_{x \not \in \gamma} \neg \rz{x}(t))
    \end{align*}

    Another property of an {\exec} is that any time the clock constraints of the state is satisfied. To do this, we need to compute the value of a clock at any time $t$ of the {\exec}. We cleverly use the clock reset predicates $R$ to find this value, and then compare it with the clock constraints of the current state. For example, if at time $t$ no transition occurs, then the value of a clock is the time elapsed since the last time it was reset. We can encode this as follows for each state $v \in V$ and clock $x \in \clocks$:
    \begin{align*}
        \forall t(\neg (\bigvee_{e \in \edges} \Tez_e(t) &\vee \Teo_e(t)) \wedge v(t) \implies \\
        &\exists r(r<t \wedge (\rz{x}(r) \vee \ro{x}(r))\\
        &\wedge \forall z(r < z < t \implies \neg (\rz{x}(z) \vee \ro{x}(z))) \\
        &\wedge \clockcons(v,x)[t - r])
    \end{align*}
    Here, $\clockcons(v,x)[t-r]$ is the clock constraint $\clockcons(v,x)$ with $x$ replaced by $t - r$. This formula says that for all time $t$, if no transition occurs at $t$ and the {\exec} is at location $v$, then there must exist a time $r$ before $t$, such that clock $x$ was reset at $r$ and for all time $z$ between $r$ and $t$, $x$ was not reset. Thus, $r$ is the last time clock $x$ was reset. Finally, we require that $t - r$ (i.e., the current value of clock $x$) satisfies the clock constraint $\clockcons(v,x)$. The satisfaction of guards for transitions can also be ensured using the same idea.
    
    A point to note here is that $\clockcons(v,x)[t-r]$ has inequalities of the form $t - r \sim c$ where $\sim \in \{<,\leq,=,\geq,>\}$. The logic $\mso$ does not have constants, so we cannot directly write formulas of the form $t-r \sim c$. However, we can still express this in the following manner:
    \begin{itemize}
        \item For any constant $c$, we will write a formula $+c(r,x)$ such that $\flow,\inter \models +c(r,x)$ iff $\inter(x) = \inter(r) + c$. For $c = 1$, we can write this directly in $\mso$ as $+1(r,x)$ 
        \item Next, using the $+1$ relation, we can write $+2$ as follows:
        $$+2(r,x) = \exists y( +1(r,y) \wedge +1(y,x)).$$
        Similarly, we can write $+3$ as: 
        $$+3(r,x) = \exists y(+2(r,y) \wedge +1(y,x)).$$
        In this manner, we can express every $+c(r,x)$ for any constant $c$ using finite sized formulas in $\mso$.
        \item Now we can rewrite $t - r \sim c$ as $t \sim r + c$. This we can write in $\mso$ as
        $$t \sim r + c \equiv \exists x(+c(r,x) \wedge t < x)$$
    \end{itemize} 
    
    In this manner, we can express all the properties of an {\exec} in $\mso$. Formally, we have the following lemma:
    \begin{lemma}
    \label{lem-tatomso}
        Given a timed automtaton $\taext$, there is a ${\mso}$ formula $\msof_{\ta}$  over free monadic predicates $\MP = V \cup T \cup R$, such that 
        \begin{enumerate}
            \item For any {\exec} $\run \in \flows{\ta}$, $\runtoflow{\ta}(\run) \models \msof_\ta$.
            \item For any flow $\flow \models \msof_{\ta}$, there exists an {\exec} $\run \in \flows{\ta}$ such that $\flow = \runtoflow{\ta}(\run)$.
        \end{enumerate}
    \end{lemma}
    \ifdefined\arxiv
        \begin{proof}
        For any flow $\flowrun: \nnreals \rightarrow \powerset{V \cup T \cup R}$ to be an {\exec}, it should satisfy the following conditions: 
    \begin{enumerate}
        \item There exists a $l \in \nnreals$ such that either:
        \begin{enumerate}
            \item for all $t \leq l$, there is exactly one $v \in V$ such that $v \in \flowrun(t)$ and for $t > l$, $\flowrun(t) = \emptyset$.
            \item for all $t < l$, there is exactly one $v \in V$ such that $v \in \flowrun(t)$ and for $t \geq l$, $\flowrun(t) = \emptyset$
        \end{enumerate}there is exactly one $v \in V$ such that $v \in \flowrun(t)$.
        \item $\flowrun(0) \cap \st \subseteq \startst$.
        \item For every $t \in \nnreals$, if $v \in \flowrun(t)$ for some $v \in V$, then there is at most one $e \in \edges$ such that $\Tez_e \in \flowrun(t)$ and there is at most one $e' \in \edges$ such that $\Teo_{e'} \in \flowrun(t)$.
        \item For any edge $e = (v_1, v_2, \guardset, \gamma)$, for all $t \in \nnreals$:
        \begin{enumerate}
            \item If $\Tez_e \in \flowrun(t)$ then, $v_2 \in \flowrun(t)$ and there exists $\epsilon > 0$ such that $t - \epsilon \geq 0$, $v_1 \in \flowrun(t')$ for all $t' \in (t - \epsilon, t)$ and $\rz{x} \in \run(t)$ iff $x \in \gamma$.
            \item $\Teo_e \in \flowrun(t)$ then, $v_1 \in \flowrun(t)$ and there exists $\epsilon > 0$ such that $v_2(t') \in \flowrun(t')$ for all $t' \in (t, t + \epsilon)$ and $\ro{x} \in \flowrun(t)$ iff $x \in \gamma$.
        \end{enumerate}
        \item For any two states $v_1 \neq v_2$, for all $t \in \nnreals$:
        \begin{enumerate}
            \item If $t \neq 0$, $v_2 \in \flowrun(t)$ and there exists $\epsilon > 0$ such that $v_1 \in \flowrun(t')$ for all $t' \in (t - \epsilon, t)$, then, there must be some edge $e = (v_1,v_2, \guardset, \gamma)$ such that $\Tez_e \in \flowrun(t)$.
            \item If $v_1 \in \flowrun(t)$ and there exists $\epsilon > 0$ such that $v_2(t') \in \flowrun(t')$ for all $t' \in (t, t + \epsilon)$, then, there must be some edge $e = (v_1,v_2, \guardset, \gamma)$ such that $\Teo_e \in \flowrun(t)$.
        \end{enumerate}
        \item For all $t \in \posreals$, if $\Tez_e \not \in \flowrun(t)$ for all $e \in \edges$, then $\rz{x} \not \in \flowrun(t)$ for all $x \in \clocks$.  Similarly, if $\Teo_e \not \in \flowrun(t)$ for all $e \in \edges$, then $\ro{x} \not \in \flowrun(t)$ for all $x \in \clocks$.
        \item For $t \in \posreals$, define a clock valuation $\clockval_t$ as follows:
        \begin{enumerate}
            \item If $\flowrun(t) \cap T = \emptyset$ or $\flowrun(t) \cap T = \{\Teo_e\}$ for some $e = (v_1,v_2,\guardset, \gamma)$, for each $x \in \clocks$, define $r_x = \sup\{t' < t \mid \rz{x} \in \run(t') \vee \ro{x} \in \run(t')\}$ (the supremum of empty set is taken to be 0). Define $\clockval_t(x) = t - r_x$.
            \item If $\{\Tez_e\} \subseteq \flowrun(t) \cap T$ for some $e = (v_1,v_2,\guardset,\gamma)$, for each $x \in \gamma$, define $\clockval_t(x) = 0$. For each $x \in X \setminus \gamma$, define $\clockval_t(x) = t - r_x$. 
        \end{enumerate}
        \item For $t = 0$, define $\clockval_t(x) = 0$ for all $x \in \clocks$.
        \item For all $t \in \nnreals$, if $\flowrun(t) \neq \emptyset$, then $\clockval_t(x) \models \beta(v,x)$ where $v \in \flowrun(t) \cap \st$.
        \item For $t \in \nnreals$ and $e = (v_1, v_2, \guardset, \gamma)$, if $\Tez_e \in \flowrun(t)$, then $\clockval_t(x) \models \guard$ for each $\guard \in \guardset \cap \conset{x}$. If $\Teo_e \in \flowrun(t)$, define $\clockval'_t(x) = 0$ if $\rz{x} \in \flowrun(t)$, else $\clockval'_t(x) = t - r_x$. Then, $\clockval_t(x) \models \guard$ for each $\guard \in \guardset \cap \conset{x}$.
     \end{enumerate}
    Furthermore, the {\exec} is accepting if this additional condition holds: 
    \begin{enumerate}
        \item[11)] Define $l = \sup\{t \mid f(t) \neq \emptyset\}$. If $f(l) \neq \emptyset$,  then $\flowrun(l) \cap V \subseteq \final$, otherwise, there must exist $\epsilon > 0$ such that for all $t \in (l-\epsilon, l)$, $\flowrun(l) \cap V \subseteq \final$.
    \end{enumerate}
    Now each of these properties can be represented in $\mso$ as follows:  
        \begin{enumerate}
            \item Property 1 can be expressed as
            \begin{align*}
        \exists l(\forall t(t \leq l \implies US(t) \wedge t > l \implies \wedge_{v \in \st} \neg v(t))) \\
        \exists l(\forall t(t < l \implies US(t) \wedge t \geq l \implies \wedge_{v \in \st} \neg v(t)))
    \end{align*} 
        ${\mso}$ does not have constants like $0$. We can have a formula $0(x)$ that is true only at $x = 0$ as follows:
        \begin{align*}
            0(x) = \forall y(x \leq y)
        \end{align*}
        \item We can ensure $\run(0) \cap \st \subseteq \startst$ using
        \begin{align*}
            \forall t(0(t) \implies \bigvee_{v \in \startst} v(t))
        \end{align*}
        \item We can ensure at most one $\Tez_e$ and $\Teo_e$ is true at any time (property 3) using
        \begin{align*}
            \forall t(&\bigvee_{v \in \st}v(t) \implies \\
            &\bigwedge_{e,e' \in \edges: e \neq e'}\neg(\Tez_e \wedge \Tez_{e'}) \wedge \neg (\Teo_e \wedge \Teo_{e'}))
        \end{align*}
        \item The consistency between the $\Tez_e$ and $v$ and $\rz{x}$ (property (4) can be ensured for each $e = (v_1,v_2,\guardset,\gamma)$ using
        \begin{align*}
            \forall t&( \Tez_e(t)\implies v_2(t) \\
            &\wedge \exists y(y < t \wedge \forall z (y < z < t \implies v_1(t))) \\
            &\wedge (\wedge_{x \in \gamma} \rz{x}(t) \wedge_{x \in \clocks \setminus \gamma} \neg \rz{x}(t)))
        \end{align*}
        Similarly we can construct a formula for $\Teo_e$.
        \item For every pair of states $v_1,v_2$, we can ensure property 5 of an {\exec} using:
        \begin{align*}
            \forall t(v_1(t) \wedge (&\exists y(t < y \wedge \forall z (t < z < y \implies  v_2(z))) \\
            &\implies \bigvee_{(v_1,v_2,\guardset, \gamma) \in \edges} \Teo_e(t)))
        \end{align*}
        Here the empty disjunct is replaced with $false$.
        \item Property 6 of an {\exec} can be ensured using
        \begin{align*}
            \forall t((\neg 0(t) \wedge \neg (\vee_{e \in \edges} \Tez_e)) \implies \neg (\vee_{x \in \clocks} \rz{x}(t)))
        \end{align*}
        \item The timing property can be ensured by observing that the truth values of $\rz{x}$ and $\ro{x}$ predicates allow us to extract the values of clocks. As an example, to ensure property (7b) of an {\exec}, for each state $v \in \st$, edge $e = (v_1,v_2, \guardset, \gamma)$ and $x \in \gamma$, we have a formula ($\clockcons(v, x)[x \rightarrow z]$ is the clock constraint $\clockcons(v, x)$ with $x$ replaced by $z$)
        \begin{align*}
            \forall t(\Tez_e(t) \implies \forall z(0(z) \implies \clockcons(v,x)[x \rightarrow z]))
        \end{align*}
        For $x \in \clocks \setminus \gamma$, we have a formula
        \begin{align*}
            \forall t(\Tez_e(t) \implies &\exists y(y < t \wedge (\rz{x}(y) \vee \ro{x}(y)) \\
            &\wedge \forall z(y < z < t \implies \neg (\rz{x}(z) \vee \ro{x}(z))) \\
            &\wedge \clockcons(v,x)[x \rightarrow (t - y)]))
        \end{align*}
        The atomic formulas in $\clockcons(v,x)[x \rightarrow (t-y)]$ are of the form $t - y \sim c$, which can be expressed in \mso{} as follows:
        \begin{align*}
            \exists z_1,z_2,\dots,z_c(&+1(y,z_1) \wedge +1(z_1,z_2) \wedge \dots \\ &\wedge +1(z_{c-1},z_c)
            \wedge t \sim z_c)
        \end{align*}
        \item The satisfaction of guards (property 10) can be handled similar to above.
        \item At $t = 0$, we set $\rz{x}$ for all clocks to 0.
        \begin{align*}
            \forall t(0(t) \implies \wedge_{x \in \clocks} \rz{x}(t))
        \end{align*}
        \item To ensure the accepting condition is satisfied, we first have a formula that is true at $x$ iff $x$ is the last transition point of the {\exec}:
        \begin{align*}
            last(x) = &(\vee_{e \in \edges}\, \Tez_e(x) \vee \Teo_e(x))\\
            &\wedge \forall y(y > x \implies \wedge_{e \in \edges}\, \neg \Tez_e(y) \wedge \neg \Teo_e(y)
        \end{align*}
        The accepting condition can be captured using:
        \begin{align*}
            \forall x (((last(x) \wedge &\neg (\vee_{e \in E} \Teo_e(x))) \implies \vee_{v \in \final} v(x)) \\
            \wedge (last(x) &\wedge (\vee_{e \in E} \Teo_e(x)) \implies \\
            &(\exists y(y > x \wedge (\vee_{v \in \final} v(y)))))
        \end{align*}
        \end{enumerate}
        The formula $\msof_{\ta}$ is the conjunction of all these formulas.

        Each of these formulas exactly capture individual properties of {\exec}s, and thus, the correctness of the construction follows.
    \end{proof}
    \else

    \fi
    The lemma also holds if we consider bounded time semantics with a time bound $N$.
    
\subsection{Decidability}
\label{sec:reduction}

We now give a reduction from the bounded time verification problem for {\hcmtl} to the satisfiability problem for $\mso$ over bounded time domains. 

Fix $N \in \posreals$ and a timed automaton $\taext$. Let $\hcmtlf$ be an \hcmtl{} formula with free path variables $\pvar$. Let the variables in $\pvar$ be ordered as $\{\pi_1,\pi_2,\dots,\pi_m\}$. Observe that the models to an \hcmtl{} formula are path environments while the models to an $\mso$ formula are flows. To reconcile this disparity, we use the encoding of {\exec}s as flows from Lemma~\ref{runtoflow}. For a path environment $\penv: \pvar \rightarrow \flowsn{\ta}$, we want to encode $\penv$ as a flow. We do this by combining the flows corresponding to each {\exec} $\penv(\pi_i)$ into one single flow and distinguishing the predicates for each {\exec} by indexing. For each path $\pi_i \in \pvar$, we define copies of the sets $\st, T$ and $R$ as $V_{i} = \{v_{i} \mid v \in V\}$, and analogously for $T_{i}$ and $R_{i}$ . Define $\MP_{i} = \{V_{i} \cup T_{i} \cup R_{i}\}$ for each $\pi_i \in \pvar$. For each path variable $\pi_i$, $\penv(\pi_i)$ is an {\exec} in $\flowsn{\ta}$ and $\runtoflow{\ta}(\penv(\pi_i))$ is a flow over the monadic predicates $\MP_i$. Define $\MP = \cup_{\pi_i \in \pvar} \MP_{i}$. We can lift the encoding $\runtoflow{\ta}$ to $\penv$ by defining a flow $\fpi{\penv}: [0,N) \rightarrow \powerset{\MP}$ as:
$$\fpi{\penv}(t) = \cup_{\pi_i \in \pvar} \runtoflow{\ta}(\penv(\pi_i))(t)$$
Our goal is to construct an $\mso$ formula $\tphi{\hcmtlf}$ that is satisfied exactly by the flows that encode path environments that satisfy $\hcmtlf$. One of the challenges in constructing $\tphi{\hcmtlf}$ is handling the last quantified path $\dagg$. Since \hcmtl{} is a logic for branching hyperproperties, the semantics of \hcmtl{} involves the last quantified path, that we represent using the variable $\dagg$. To construct $\tphi{\hcmtlf}$, we need to encode the last quantified path into $\tphi{\hcmtlf}$. One option is to encode $\dagg$ as a first order variable in $\tphi{\hcmtlf}$ that takes values in $\{1,2,\dots,m\}$. However, this does not work because by the semantics of $\mso$, $\dagg$ can only take values in $[0,N)$. Hence, if the number of free path variables, $m$ is larger than $N$, it will not be possible to cover all values of $\dagg$.

    We overcome this by instead having $m+1$ formulas indexed $\tphi{\hcmtlf}_i$ for $i \in \{0,1,2,\dots,m\}$. And the property that we preserve is the following: for any path environment $\penv$ and $t \in [0,N)$, $$\hcmtlmodelsargbdd{\penv}{t}{\pi_i} \hcmtlf \text{ iff } \fpi{\penv} \models \tphi{\hcmtlf}_i(t).$$
    If $\dagg = \epsilon$, then the property we preserve is
    $$\hcmtlmodelsargbdd{\penv}{0}{\epsilon} \hcmtlf \text{ iff } \fpi{\penv} \models \tphi{\hcmtlf}_0(0).$$
    We do this by constructing the formulas $\tphi{\hcmtlf}_i$ inductively. Translating quantifier free formulas to $\mso$ is straightforward and is identical for all $i \in \{0,1,2,\dots, m\}$. For example, if $\hcmtlf = p_{\pi_j}$, this means that at time $t$, $\pi_j$ is in a state where the proposition $p$ is true. The corresponding $\mso$ formula is:
    $$\tphi{\hcmtlf}_i(x) = \bigvee_{v \in V: p \in \stlab(v)} v_j(x)$$
    Another example is $\hcmtlf = \hcmtlf_1 U_I \hcmtlf_2$. This formula says that at some future time $y > t$, such that $y \in t + I$, the formula $\hcmtlf_2$ must hold, and for all time $t < z < y$, the formula $\hcmtlf_1$ holds. This naturally translates to $\mso$ as
    \begin{align*}
        \tphi{\hcmtlf}_i(x) = \exists y(&x < y \wedge \tphi{\hcmtlf_2}_i(y) \wedge \\
        &\forall z (x < z < y \implies \tphi{\hcmtlf_1}_i(z)) \wedge y - x \in I)
    \end{align*} 

    The construction for the $\exists$ and $\forall$ quantifiers involves the translation of timed automata to $\mso$ that we described in lemma \ref{lem-tatomso}. Suppose $\hcmtlf = \exists \pi_{m+1} \hcmtlf_1$. The corresponding $\mso$ formula $\tphi{\hcmtlf}_i$ should read `there exists an {\exec} that is identical to $\pi_i$ up to the current time, and the path environment $\penv$ augmented with this new {\exec} satisfies $\tphi{\hcmtlf_1}_{m+1}$.' Since we are quantifying over {\exec}s of the automaton, in $\tphi{\hcmtlf}$, we quantify over flows that satisfy the formula $\msof_\ta$ from Lemma~\ref{lem-tatomso}, and this can only be done in a second order logic. The formula $\tphi{\hcmtlf}_i$ is constructed as:
    \begin{align*}
        \tphi{\hcmtlf}_i(x) = \exists &
        V_{m+1}, T_{m+1}, R_{m+1} \\
        &(\msof_\ta(V_{m+1},T_{m+1},R_{m+1}) \\
        &\wedge \forall y(0 \leq y \leq x \implies (\MP_i(y) \iff \MP_{m+1}(y))) \\
        & \wedge \tphi{\hcmtlf_1}_{m+1}(x))
            \end{align*}
    Here $\MP_i(y) \iff \MP_{m+1}(y)$ is an abbreviation for 
    $$\wedge_{p \in \MP} (p_i(y) \iff p_{m+1}(y))$$ i.e. the $i$th copy and $m+1$st copy of the predicates have the same truth values at $y$.
    
    Putting these ideas together, we have the following lemma.
    \begin{lemma}
        \label{lem-hcmtl-mso}
        Fix $N \in \posreals{}$, an \hcmtl{} formula $\hcmtlf$ and a timed automaton $\taext$. There exist $m+1$ $\mso$ formulas $\tphi{\hcmtlf}_0(x), \tphi{\hcmtlf}_1(x),$ $ \dots, \tphi{\hcmtlf}_m(x)$, each over $\MP$ with one free first order variable $x$ such that, for $t \in [0,N)$ and a path environment $\penv$, 
        \begin{enumerate}
            \item For $i \in \{1,2,\dots,m\}$, $\hcmtlmodelsargbdd{\penv}{t}{\pi_i} \hcmtlf$ iff $\fpi{\penv} \models \tphi{\hcmtlf}_i(t)$
        \item $\hcmtlmodelsargbdd{\penv}{0}{\epsilon} \hcmtlf$ iff $\fpi{\penv} \models \tphi{\hcmtlf}_0(0)$. \end{enumerate}
    \end{lemma}
    \ifdefined\arxiv
    \begin{proof}
        We construct $\tphi{\hcmtlf}_0, \tphi{\hcmtlf}_1, \dots, \tphi{\hcmtlf}_m$ by inducting on the structure of $\hcmtlf$.

        Base Cases:
        \begin{enumerate}
            \item $\hcmtlf = p_{\pi_j}$. In this case, $\tphi{\hcmtlf}_i(x) = \vee_{v \in V: p \in \alpha(v)}v_j(x)$ for all $i \in \{0,1,2,\dots,m\}$.
            \item $\hcmtlf = \neg p_{\pi_j}$. In this case, $\tphi{\hcmtlf}_i(x) = \vee_{v \in V: p \not \in \alpha(v)}v_j(x)$ for all $i \in \{0,1,2,\dots,m\}$.
        \end{enumerate}
        Inductive Cases:
        \begin{enumerate}
            \item $\hcmtlf = \hcmtlf_1 \vee \hcmtlf_2$. In this case, $\tphi{\hcmtlf}_i(x) = \tphi{\hcmtlf_1}_i(x) \vee \tphi{\hcmtlf_2}_i(x)$ for all $i \in \{0,1,2,\dots,m\}$.
            \item $\hcmtlf = \hcmtlf_1 \wedge \hcmtlf_2$. In this case, $\tphi{\hcmtlf}_i(x) = \tphi{\hcmtlf_1}_i(x) \wedge \tphi{\hcmtlf_2}_i(x)$ for all $i \in \{0,1,2,\dots,m\}$.
            \item $\hcmtlf = F_I \hcmtlf_1$. In this case,
            $$\tphi{\hcmtlf}(x) = \exists y(x < y \wedge y - x \in I \wedge \tphi{\hcmtlf_1}(y)).$$
            \item $\hcmtlf = G_I \hcmtlf_1$. In this case,
            $$\tphi{\hcmtlf}(x) = \forall y (x < y \wedge x - y \in I \implies \tphi{\hcmtlf_1}(y))$$
            \item $\hcmtlf = \hcmtlf_1U_I\hcmtlf_2$. In this case, for all $i \in \{0,1,\dots,m\}$
            \begin{align*}
                \tphi{\hcmtlf}_i(x) = \exists y(&x < y \wedge \tphi{\hcmtlf_2}_i(y) \\
                &\wedge \forall z(x < z < y \implies \tphi{\hcmtlf_1}_i(z)) \wedge y - x \in I)
            \end{align*}
            \item $\hcmtlf = \exists \pi_{m+1} \hcmtlf_1$. In this case, for all $i \in \{1,2,\dots, m\}$
            \begin{align*}
                \tphi{\hcmtlf}_i(x) = \exists &
                V_{m+1}, T_{m+1}, R_{m+1}, \top_{m+1} \\
                &(\msof_\ta(V_{m+1},T_{m+1},R_{m+1},\top_{m+1}) \\
                &\wedge \forall y(0 \leq y \leq x \implies (\MP_i(y) \iff \MP_{m+1}(y))) \\
                &\wedge \tphi{\hcmtlf_1}_{m+1}(x))
            \end{align*}
            here $\MP_i(y) \iff \MP_{m+1}(y)$ is an abbreviation for $\wedge_{p \in \MP}(p_i(y) \iff p_{m+1}(y))$. $\exists V_{m+1}$ is an abbreviation for $\exists_{v \in V} v_{m+1}$ and similarly for $\exists T_{m+1}$ and $\exists R_{m+1}$. 
            
            For $i = 0$, we have:
            \begin{align*}
                \tphi{\hcmtlf}_0(x) = \exists &
                V_{m+1}, T_{m+1}, R_{m+1}, \top_{m+1}\\
                &(\msof_\ta(V_{m+1},T_{m+1},R_{m+1},\top_{m+1}) \\
                &\wedge \tphi{\hcmtlf_1}_{m+1}(x))
            \end{align*}
            \item $\hcmtlf = \forall \pi_{m+1} \hcmtlf_1$. In this case, for all $i \in \{1,2,\dots, m\}$
            \begin{align*}
                \tphi{\hcmtlf}_i(x) = \forall 
                V_{m+1}, T_{m+1}, &R_{m+1}, \top_{m+1} \\
                (\msof_\ta(V_{m+1},T_{m+1},&R_{m+1},\top_{m+1}) \\
                \wedge \forall y(0 \leq y \leq x &\implies \\
                (\MP_i(y) &\iff \MP_{m+1}(y))) \\
             \wedge \tphi{\hcmtlf_1}_{m+1}&(x))
            \end{align*}
            For $i = 0$, we have:
            \begin{align*}
                \tphi{\hcmtlf}_0(x) = \forall &
                V_{m+1}, T_{m+1}, R_{m+1}, \top_{m+1}\\
                &(\msof_\ta(V_{m+1},T_{m+1},R_{m+1},\top_{m+1}) \\
                &\wedge \tphi{\hcmtlf_1}_{m+1}(x))
            \end{align*}
        \end{enumerate}
        We will prove correctness of the construction by inducting on the structure of $\hcmtlf$. Pick any path environment $\penv$ and $t \in [0,N)$. First we will prove that for $i \in \{1,2,\dots,m\}, \hcmtlmodelsarg{\penv}{t}{\pi_i} \hcmtlf$ iff $\fpi{\penv} \models\tphi{\hcmtlf}_i(t)$. Base Cases:
        \begin{enumerate}
            \item $\hcmtlf = p_{\pi_j}$. From the semantics of \hcmtl{}, $\hcmtlmodelsarg{\penv}{t}{\pi_i} p_{\pi_j}$ iff $t \in \penv(\pi_j)$ and $v \in \penv(\pi_j)(t)$ for some $v \in V$ with $p \in \stlab(v)$. By the definition of extension of an {\exec}, this happens iff $v_j \in \fpi{\penv}(t)$. Finally, $v_j \in \fpi{\penv}(t)$ iff $\fpi{\penv} \models \tphi{\hcmtlf}_i(t)$. Again, this argument works for any $i \in \{1,2,\dots,m\}$.
            \item $\hcmtlf = \neg p_{\pi_j}$. From the semantics of \hcmtl{}, $\hcmtlmodelsarg{\penv}{t}{\pi_i} p_{\pi_j}$ iff $t \in \penv(\pi_j)$ and $v \in \penv(\pi_j)(t)$ for some $v \in V$ with $p \not \in \stlab(v)$. By the definition of extension of an {\exec}, this happens iff $v_j \in \fpi{\penv}(t)$. Finally, $v_j \in \fpi{\penv}(t)$ iff $\fpi{\penv} \models \tphi{\hcmtlf}_i(t)$. Again, this argument works for any $i \in \{1,2,\dots,m\}$.
        \end{enumerate}
        Inductive Cases:
        \begin{enumerate}
            \item $\hcmtlf = \hcmtlf_1 \vee \hcmtlf_2$. From \hcmtl{} semantics, $\hcmtlmodelsarg{\penv}{t}{\pi_i} \hcmtlf_1 \vee \hcmtlf_2$ iff $\hcmtlmodelsarg{\penv}{t}{\pi_i} \hcmtlf_1$ or $\hcmtlmodelsarg{\penv}{t}{\pi_i} \hcmtlf_2$. By induction hypothesis, this happens iff $\fpi{\penv} \models \tphi{\hcmtlf_1}_i(t)$ or $\fpi{\penv} \models \tphi{\hcmtlf_2}_i(t)$, iff $\fpi{\penv} \models \tphi{\hcmtlf_1}_i(t) \vee \tphi{\hcmtlf_2}_i(t)$. This argument works for all $i \in \{1,2,\dots,m\}$.
            \item The argument for $\hcmtlf = \hcmtlf_1 \wedge \hcmtlf_2$ is exactly the same as the previous case.
            \item The proof for $F_I$ and $G_I$ is similar to $U_I$, so we prove only for $U_I$.
            \item $\hcmtlf = \hcmtlf_1U_I\hcmtlf_2$ and
            \begin{align*}
                \tphi{\hcmtlf}_i(x) = \exists y(&x < y \wedge \tphi{\hcmtlf_2}_i(y) \\
                &\wedge \forall z(x < z < y \implies \tphi{\hcmtlf_1}_i(z)) \wedge y - x \in I)
            \end{align*}
            By \hcmtl{} semantics, $\hcmtlmodelsarg{\penv}{t}{\pi_i} \hcmtlf_1 U_I \hcmtlf_2$ iff there is some $t' > t$ such that $\hcmtlmodelsarg{\penv}{t'}{\pi_i} \hcmtlf_2$ and for all $t < t'' < t'$, $\hcmtlmodelsarg{\penv}{t''}{\pi_i} \hcmtlf_1$ and $t' - t \in I$. By induction hypothesis, this happens iff
            \begin{align*}
                &\fpi{\penv} \models \tphi{\hcmtlf_2}(t') \\
                &\fpi{\penv} \models \forall z (t < z < t' \implies \tphi{\hcmtlf_1}_i(z)) \\
                &\fpi{\penv} \models t < t' \\
                &\fpi{\penv} \models t' - t \in I
            \end{align*}
            Now, by the semantics of first order existential quantification of $\mso$, this happens iff $\fpi{\penv} \models \tphi{\hcmtlf}_i(t)$. This argument works for any $i \in \{1,2,\dots,m\}$.
            \item $\hcmtlf = \exists \pi_{m+1} \hcmtlf_1$ and
            \begin{align*}
                \tphi{\hcmtlf}_i(x) = \exists 
                V_{m+1}, T_{m+1}, &R_{m+1}, \top_{m+1} \\
                (\msof_\ta(V_{m+1},T_{m+1},&R_{m+1},\top_{m+1}) \\
                \wedge \forall y(0 \leq y \leq x &\implies\\
                (\MP_i(y) &\iff \MP_{m+1}(y))) \\
                \wedge \top_i(x) \wedge \tphi{\hcmtlf_1}_{m+1}&(x))
            \end{align*}
            Suppose $\hcmtlmodelsarg{\penv}{t}{\pi_i} \hcmtlf$. Then, $t \leq \len{\penv(\pi_i)}$  and there is an {\exec} $g$ in $\flowsn{\ta}$ such that $g(t') = \pi_i(t')$ for all $0 \leq t' \leq t$ and the path environment $\penv'(\pi_{m+1}) = g$ and $\penv'(\pi_j) = \penv(\pi_j)$ for all $j \in \{1,2,\dots,m\}$ satisfies $\hcmtlmodelsarg{\penv'}{t}{\pi_{m+1}} \hcmtlf_1$. By induction hypothesis, the flow $\fpi{\penv'} \models \tphi{\hcmtlf_1}_{m+1}(t)$. Since, $g(t') = \pi_i(t')$ for all $0 \leq t' \leq t$,
            $$\fpi{\penv'} \models \forall y(0 \leq y \leq x \implies (\MP_i(y) \iff \MP_{m+1}(y))). $$
            Also, since $g \in \flowsn{\ta}$, by lemma \ref{lem-tatomso}, $$\runtoflow{\ta}(g) \models \msof_{\ta}(V_{m+1}, T_{m+1}, R_{m+1}, \top_{m+1}).$$
            Since $\fpi{\penv'}$ restricted to the monadic predicates $\MP_{m+1}$ is $\runtoflow{\ta}(g)$, we have $$\fpi{\penv'} \models \msof_{\ta}(V_{m+1}, T_{m+1}, R_{m+1}, \top_{m+1}).$$
            Thus, by the second order existential quantification semantics of $\mso$, we get $\fpi{\penv} \models \tphi{\hcmtlf}_i(t)$. Conversely, suppose $\fpi{\penv} \models \tphi{\hcmtlf}_i(t)$. Then, there is a flow $h: [0,N) \rightarrow \MP_{m+1}$ such that $h \models \msof_{\ta}(V_{m+1}, T_{m+1}, R_{m+1}, \top_{m+1}$). By lemma \ref{lem-tatomso}, there is an {\exec} $g \in \flowsn{\ta}$ such that $h = \runtoflow{\ta}(g)$ and
            \begin{align*}
                g \models &\forall y(0 \leq y \leq t \implies (\MP_i(y) \iff \MP_{m+1}(y))).
            \end{align*}
            Let $\penv'$ be the path environment defined as $\penv'(\pi_{m+1}) =g$ and $\penv'(\pi_{j}) = \penv(\pi_j)$ for $j \in \{1,2,\dots, m\}$. Observe that $h = \fpi{\penv'}$ satisfies $h \models \tphi{\hcmtlf}_{m+1}(t)$. Thus, by the induction hypothesis, $\hcmtlmodelsarg{\penv'}{t}{\pi_{m+1}} \hcmtlf_1$ and thus, by existential semantics of \hcmtl{}, $\hcmtlmodelsarg{\penv}{t}{\pi_i} \hcmtlf$. Again, this argument works for all $i \in \{1,2,\dots,m\}$.
            \item The argument for universal quantification is similar to the argument for existential quantification.
        \end{enumerate}

        For the case of $i = 0$, the arguments differ only for the existential and universal quantification, which we present here.
        \begin{enumerate}
            \item $\hcmtlf = \exists \pi_{m+1} \hcmtlf_1$ and
            \begin{align*}
                \tphi{\hcmtlf}_0(x) = \exists &
                V_{m+1}, T_{m+1}, R_{m+1}, \top_{m+1}\\
                &(\msof_\ta(V_{m+1},T_{m+1},R_{m+1},\top_{m+1}) \\
                &\wedge \tphi{\hcmtlf_1}_{m+1}(x))
            \end{align*}
            By \hcmtl{} semantics, $\hcmtlmodelsarg{\penv}{0}{\epsilon} \hcmtlf$ iff exists an {\exec} $g \in \flowsn{\ta}$ such that $\hcmtlmodelsarg{\penv'}{0}{\pi_{m+1}} \hcmtlf_1$. By lemma \ref{lem-tatomso}, $\runtoflow{\ta}(g) \models \msof_\ta(V_{m+1},T_{m+1},R_{m+1},\top_{m+1})$, and by induction hypothesis, $\fpi{\penv'} \models \tphi{\hcmtlf_1}_{m+1}(0)$. Since $\fpi{\penv'}$ is extending $\fpi{\penv}$ by $\runtoflow{\ta}(g)$, we have $\fpi{\penv} \models \tphi{\hcmtlf}_0(0)$.
            \item The argument for universal quantification is similar to the one for existential quantification.
        \end{enumerate}
    \end{proof}
    \else

    \fi
    We thus have this corollary
    \begin{corollary}
        For an \hcmtl{} sentence $\hcmtlf$, $\hcmtlmodelsargbdd{\emptypenv}{0}{\epsilon} \hcmtlf$ iff $\tphi{\hcmtlf}_0(0)$ is a valid $\mso$ sentence over $\tdomain = [0,N)$.
    \end{corollary}
    Thus, by the corollary, time bounded verification of an \hcmtl{} sentence $\hcmtlf$ in the interval-based semantics reduces to checking satisfiability of the $\mso$ formula $\tphi{\hcmtlf}_0(0)$ over the bounded time domain $[0,N)$, which is known to be decidable~\cite{boundedtime}. Thus, we get our main result
    \begin{theorem}
        \label{thm-decidability}
        Bounded Time verification problem of \hcmtl{} is decidable in the interval-based semantics.
    \end{theorem}
    \begin{remark}
        In our presentation, we have restricted clock constraints in timed automata to allow only comparisons with natural numbers, and the intervals $I$ in temporal operators such as $U_I$ to have natural number end points. This is done for the sake of simplicity and all the results presented here carry over if we allow non-negative rational numbers instead of natural numbers. In case of rational bounds, the model checking problem can be reduced to a model checking problem with only natural number bounds by appropriately scaling all constants appearing the the timed automata and the \hcmtl{} formula. The appropriate scaling factor will be the least common multiple of all denominators occurring in all the rational constants.
    \end{remark}
    \section{Point-based Semantics}
    \label{sec:ptsemantics}
    We define a point-based semantics for our logic \hcmtl{}. We first give a point-based semantics for timed automata, then move on to \hcmtl{}, and finally present our results for the point based semantics.
    \subsection{Timed Automata}
    In the interval-based semantics, the system is under observation at all times. On the other hand, in the point-based semantics, the system is observed at discrete time points when events (marked by propositions that are true at the event) occur. A timed automata in the point based semantics over a set of propositions $\props$ is a tuple $\pttaext$ where
    \begin{itemize}
        \item $\ptst$ is a finite set of states.
        \item $\ptstartst \in \ptst$ is the start state. 
        \item $\ptclocks$ is a finite set of clocks.
        \item $\ptedges \subseteq \ptst \times \powerset{\props} \times \powerset{\conset{\clocks}} \times \powerset{\clocks} \times \ptst$ is the transition relation. A transition $e = (s_1, \event, \guardset, \gamma, s_2)$ is a transition from state $s_1$ to $s_2$ on event $\event$ that satisfies the guard $\guardset$ and resets the clocks in $\gamma$.
        \item $\ptfinal \subseteq \ptst$ is the set of final states.
    \end{itemize}
    For a transition $e = (s_1,\event, \guardset, \gamma, s_2)$, we will call $\sigma$ the event labelling $e$. An {\exec} of $\ptta$ is a finite sequence 
    $$\ptrun = (s_0, \clockval_0) \xrightarrow{e_1,\evtime_1}(s_1,\clockval_1) \xrightarrow{e_2,\evtime_2}\dots \xrightarrow{e_n,\evtime_n}(s_n,\clockval_n)$$ 
    where $s_i \in \ptst, e_i \in \ptedges$ for all $i \geq 1$ and $\clockval_i$ is a clock valuation for each $i$ such that the following hold
    \begin{itemize}
        \item $\clockval_0(x) = 0$ for all $x \in \ptclocks$
        \item For each $i \in \{1, 2, \dots, n\}$,  $e_i = (s_{i-1}, \event_i, \guardset_i, \gamma_i, s_{i}) \in \ptedges$ for some $\event_i, \guardset_i$, and $\gamma_i$.
        \item For $i \geq 1$, $\clockval_{i}(x) = \clockval_{i-1}(x) + (\evtime_{i} - \evtime_{i-1})$ if $x \not \in \gamma_i$ and $\clockval_{i+1}(x) = 0$ if $x \in \gamma_i$.
        \item Finally, $\clockval_{i-1} + (\evtime_{i} - \evtime_{i-1}) \models \cons$ for every $\cons \in \guardset_i$ ($t_0$ is defined to be 0).
    \end{itemize}
     The {\exec} is said to be accepting if $s_n \in F$. The duration of the {\exec}, denoted $\len{\ptrun}$, is defined to be $t_n$. Define $\ptruns{\ptta}$ to be the set of all \emph{accepting} {\exec}s of $\ptta$. For an {\exec} $\ptrun$, and $t \in \nnreals$, we will say $t \in \ptrun$ iff $t = \evtime_i$ for some $i$. For $t \in \ptrun$, where $t = t_i$ define $\event_{\ptrun}(t) = \event_i$. For $t \leq \len{\ptrun}$, let $t_i = \sup \setpred{x \in \ptrun}{x \leq t}$. We will say an {\exec}
    \begin{align*}
        \ptrun' = (s_0, \clockval_0) \xrightarrow{e_1,\evtime_1}&(s_1,\clockval_1)\xrightarrow{e_2,\evtime_2}\dots\xrightarrow{e_i, \evtime_i} (s_i, \clockval_i) \xrightarrow{e'_{i+1},\evtime'_{i+1}}
        \\&(s'_{i+1},\clockval'_{i+1})\xrightarrow{e'_{i+1},\evtime'_{i+1}} \dots \xrightarrow{e'_m,\evtime'_m}(s_m,\clockval_m)
    \end{align*}
    with $\evtime'_{i+1} > t$ is an extension of $\ptrun$ from $t$.
    \subsection{{\hcmtl} in Point-Based Semantics}
    Given an \hcmtl{} formula $\hcmtlf$ with free path variables $\pvar$ and a timed automaton in the point-based semantics $\ptta$, a path environment is a map $\ptpenv: \pvar \rightarrow \ptruns{\ptta}$. We will say $t \in \nnreals$ is an event point in $\ptpenv$ if $t \in \ptpenv(\pi)$ for some $\pi \in \pvar$, and we denote this by $t \in \ptpenv$. For $t \in \nnreals$, and $\dagg$ taking values in $\pvar \cup \{\epsilon\}$, the satisfaction relation $\pthcmtlargta{\ptpenv}{t}{\dagg}{\ptta} \hcmtlf$ is defined inductively as follows:
    \begin{itemize}
        \item $\pthcmtlmodels p_\pi$ iff $t \in \ptpenv(\pi)$ and $p \in \event_{\ptpenv(\pi)}(t)$.
        \item $\pthcmtlmodels \neg p_\pi$ iff $t \in \ptpenv(\pi)$ and $p \not \in \event_{\ptpenv(\pi)}(t)$.
        \item $\pthcmtlmodels \hcmtlf_1 \vee \hcmtlf_2$ iff $\pthcmtlmodels \hcmtlf_1$ or $\pthcmtlmodels \hcmtlf_2$.
        \item $\pthcmtlmodels \hcmtlf_1 \wedge \hcmtlf_2$ iff $\pthcmtlmodels \hcmtlf_1$ and $\pthcmtlmodels \hcmtlf_2$.
        \item $\pthcmtlmodels F_I \hcmtlf$ iff there exists $t' > t$ such that $t' - t \in I$, $t' \in \ptpenv$ and $\pthcmtlargta{\ptpenv}{t'}{\dagg}{\ptta} \hcmtlf$.
        \item $\pthcmtlmodels G_I \hcmtlf$ iff for all $t' > t$ such that $t' - t \in I$ and $t' \in \ptpenv$,  $\pthcmtlargta{\ptpenv}{t'}{\dagg}{\ptta} \hcmtlf$.
        \item $\pthcmtlmodels \hcmtlf_1 U_I \hcmtlf_2$ iff there exists $t' > t$ such that $t' - t \in I$, $t' \in \ptpenv$, \ $\pthcmtlargta{\ptpenv}{t'}{\dagg}{\ptta} \hcmtlf_2$; and $\pthcmtlargta{\ptpenv}{t''}{\dagg}{\ptta} \hcmtlf_1$  for all $t < t'' < t'$ such that $t'' \in \ptpenv$. 
        \item $\pthcmtlmodels \exists \pi \hcmtlf$ iff there is an {\exec} $\ptrun \in \ptruns{\ptta}$ such that $\pthcmtlargta{\ptpenv[\pi \mapsto \ptrun]}{t}{\pi}{\ptta} \hcmtlf$ and either (a) $\dagg = \epsilon$ and $t=0$, or (b) $t \leq \len{\ptpenv(\dagg)}$ and $\ptrun$ is an extension of $\ptpenv(\dagg)$ from $t$.
        \rmv{
        \begin{itemize}
            \item If $\dagg \in \pvar$ and $t \leq \len{\ptpenv(\dagg)}$, then there exists an {\exec} $\ptrun$ in $\ptruns{\ptta}$ that is an extension of $\ptpenv(\dagg)$ from $t$. If $\dagg = \epsilon$ and $t = 0$, then there exists an {\exec} $\ptrun$ in $\ptruns{\ptta}$.
            \item Define a path environment $\ptpenv'$ as $\ptpenv'(\pi) = \ptrun$ and $\ptpenv'(\pi') = \ptpenv(\pi')$ for all $\pi' \neq \pi$.
            \item $\pthcmtlargta{\ptpenv'}{t}{\pi}{\ptta} \hcmtlf$.
        \end{itemize}
        }
        \item $\pthcmtlmodels \forall \pi \hcmtlf$ iff for every $\ptrun \in \ptruns{\ptta}$, if either (a) $\dagg = \epsilon$ and $t = 0$, or (b) $t \leq \len{\ptpenv(\dagg)}$ and $\ptrun$ is an extension of $\ptpenv(\dagg)$ from $t$, then $\pthcmtlargta{\ptpenv[\pi \mapsto \ptrun]}{t}{\pi}{\ptta} \hcmtlf$.
        \rmv{
        \begin{itemize}
            \item If $\dagg \in \pvar$ and $t \leq \len{\ptpenv(\dagg)}$, then for all {\exec}s $\ptrun$ in $\ptruns{\ptta}$ that is an extension of $\ptpenv(\dagg)$ from $t$. If $\dagg = \epsilon$ and $t = 0$, then for all {\exec}s $\ptrun$ in $\ptruns{\ptta}$.
            \item Define a path environment $\ptpenv'$ as $\ptpenv'(\pi) = \ptrun$ and $\ptpenv'(\pi') = \ptpenv(\pi')$ for all $\pi' \neq \pi$.
            \item $\pthcmtlargta{\ptpenv'}{t}{\pi}{\ptta} \hcmtlf$.
        \end{itemize}
        }
    \end{itemize}
    We can define time bounded semantics for a time bound $N \in \posreals$ in a manner similar as we did for the interval-based semantics in Section~\ref{sec:semantics}.
    \subsection{Point-Based vs Interval-Based Semantics}
    The first observation we make between the two semantics is that timed automata in the interval-based semantics can simulate timed automata in the point-based semantics. The idea is that an {\exec} of the automaton $\ptta$ 
    $$\ptrun = (s_0, \clockval_0) \xrightarrow{e_1,\evtime_1}(s_1,\clockval_1) \xrightarrow{e_2,\evtime_2}\dots \xrightarrow{e_n,\evtime_n}(s_n,\clockval_n)$$
    can be thought of in the interval semantics as the {\exec} 
    \begin{align*}
        \run = &(s_0, [0,t_1)),(e_1, [t_1,t_1]),(s_1,(t_1,t_2)),(e_2, [t_2,t_2]), \\
        &\dots,(s_{n-1},(t_{n-1},t_n),(e_n,[t_n,t_n]).
    \end{align*}
    where transitions are marked by singular intervals. Here, if $t_1 = 0$, the {\exec} starts at $(e_1, [t_1,t_1])$. To achieve this, we construct an automaton in the interval-based semantics, we will call it $\taenc{\ptta}$,  that has as its states all the states and edges of $\ptta$. $\taenc{\ptta}$ alternates between states that correspond to states of $\ptta$ and states that correspond to edges of $\ptta$. To ensure that this automaton remains in the edge states only for singular intervals, we have a special clock $\xsing$. This clock gets reset whenever we enter a state corresponding to an edge, and that state has the constraint $\xsing = 0$. In $\taenc{\ptta}$, the states corresponding to edges have the clock constraints of the edge to ensure the timing constraints of $\ptrun$ are satisfied.
    
    Given a timed automaton $\pttaext$ in the point-based semantics, define a timed automaton $\taenc{\ptta}$ over the set of propositions $\props \cup \ptst \cup \{\marker\}$ in the interval-based semantics as, $\taenc{\ptta} = (\st, \startst, \stlab, X', \clockcons, \edges, \final)$ where:
    \begin{itemize}
        \item $\st = \ptst \cup \{e \mid e \in \ptedges\}$.
        \item $\startst = \{s_0\} \cup \{e \in \ptedges \mid e = (s_0, \event, \cons, \gamma, s')\}$.
        \item $\stlab(s) = \{s\}$ for all $s \in \ptst$ and $\stlab(e) = \event \cup \{\marker\}$ for $e = (s_1,\event, \guardset, \gamma, s_2)$.
        \item $X' = X \cup \{\xsing\}$.
        \item $\clockcons(s,x) = \text{true}$ for all $s \in S$ and $x \in X'$. For $e = (s_1,\event, \guardset, \gamma, s_2)$, $\clockcons(e,x) = \wedge_{\cons \in G_x} \cons$ where $G_x = \guardset \cap \conset{x}$  for $x \in \clocks$, and  $\clockcons(e,\xsing) = \{\xsing = 0\}$.
        \item $\edges = \{(s, e, \set{\text{true}}, \{\xsing\}) \mid e = (s,\event, \guardset, \gamma, s') \in \ptedges \} \cup \{(e, s', \set{\text{true}}, \gamma) \mid e = (s,\event, \guardset, \gamma, s') \in \ptedges\}$.
        \item $\final = \{e \in \ptedges \mid e = (s_1, \event, \guardset, \gamma, s_2) \text{ and } s_2 \in \ptfinal\}$.
    \end{itemize}
     The $\marker$ proposition is used to mark points where transitions occur and is used later to reduce the verification problem in the point-based semantics to the verification problem in the interval-based semantics. Consider the map $\enc: \ptruns{\ptta} \rightarrow \flows{\taenc{\ptta}}$ that maps an {\exec} 
    $$\ptrun = (s_0, \clockval_0) \xrightarrow{e_1,\evtime_1}(s_1,\clockval_1) \xrightarrow{e_2,\evtime_2}\dots \xrightarrow{e_n,\evtime_n}(s_n,\clockval_n)$$
    of $\ptta$ to the {\exec} 
    \begin{align*}
        \run = &(s_0, [0,t_1)),(e_1, [t_1,t_1]),(s_1,(t_1,t_2)),(e_2, [t_2,t_2]), \\
        &\dots,(s_{n-1},(t_{n-1},t_n),(e_n,[t_n,t_n])
    \end{align*}
    of $\taenc{\ptta}$. The automaton $\taenc{\ptta}$ simulates exactly the set of accepting {\exec}s of $\ptta$ when transformed using the map $\enc$. Thus, we have the following lemma.
    \begin{lemma}
        \label{lem-bijec}
        $\enc$ is a bijection from $\ptruns{\ptta}$ to $\flows{\taenc{\ptta}}$.
    \end{lemma}
    \ifdefined\arxiv
    \begin{proof}
        First we prove that for any {\exec} $\ptrun \in \ptruns{\ptta}$, $\enc(\ptrun) \in \flows{\taenc{\ptta}}$ i.e it is an accepting {\exec} of $\taenc{\ptta}$. Let 
        $$\ptrun = (s_0, \clockval_0) \xrightarrow{e_1,\evtime_1}(s_1,\clockval_1) \xrightarrow{e_2,\evtime_2}\dots \xrightarrow{e_n,\evtime_n}(s_n,\clockval_n).$$
        Then, 
        \begin{align*}
        \enc(\ptrun) = &(s_0, [0,t_1)),(e_1, [t_1,t_1]),(s_1,(t_1,t_2)),(e_2, [t_2,t_2]), \\
        &\dots,(s_{n-1},(t_{n-1},t_n),(e_n,[t_n,t_n]).
    \end{align*}
    Since $e_i = (s_{i-1}, \sigma_i, \cons_i,\gamma_i, s_i) \in \ptedges$ for each $i \in \{1,2,\dots,n\}$, by definition of $\taenc{\ptta}$, we have, $(s_{i-1},e_i,True, \{\xsing\}) \in  \edges$ and $(e_i,s_i,True, \gamma_i) \in \edges$. Thus, the transitions in {\exec} $\enc(\ptrun)$ are legitimate transitions. Now we need to argue that the clock constraints are always satisfied. For any state $s_i$ and clock $x \in X'$, the clock constraint is $True$ and thus it is trivially satisfied. Similarly, the guards in edges of $\taenc{\ptta}$ are trivially satisfied. Observe that for any $i \in \{0,1,2,\dots, n\}$, the clock valuation at time $t_i$ in run $\enc(\ptrun)$ is $\clockval_{i-1}(x) + (t_{i} - t_{i-1})$ for $x \in \clocks$ and $0$ for $\xsing$ ($t_0$ is defined to be 0). Since $\ptrun$ is an {\exec} of $\ptta$, $\clockval_{i-1} + (t_i - t_{i-1}) \models \cons_i$ and thus, $\clockval_{i-1}(x) + (t_{i} - t_{i-1}) \models \clockcons(e_i,x)$ for all $x \in \clocks$. $\clockcons(e_i,\xsing) = \{\xsing = 0\}$ and $\xsing$ has value $0$ at time $t_i$. Thus, all clock constraints are satisfied at time $t_i$ and this is true for all $i \in \{1,2,\dots,n\}$. Thus, $\enc(\ptrun) \in \flows{\taenc{\ptta}}$. The map is clearly injective. To show it is surjective, pick any {\exec} $\run \in \flows{\taenc{\ptta}}$ 
    \begin{align*}
        \run = &(s_0, [0,t_1)),(e_1, [t_1,t_1]),(s_1,(t_1,t_2)),(e_2, [t_2,t_2]), \\
        &\dots,(s_{n-1},(t_{n-1},t_n),(e_n,[t_n,t_n]).
    \end{align*}
    Define clock valuations over $\clocks$ as:
    \begin{itemize}
        \item $\clockval_0(x) = 0$ for all $x \in \clocks$.
        \item Let $e_i = (s_{i-1}, \sigma_i, \cons_i, \gamma_i, s_i)$. $\clockval_i(x) = \clockval_{i-1}(x) + (t_i - t_{i-1})$ for $x \not \in \gamma_i$ and $\clockval_i(x) = 0$ for $x \in \gamma_i$ ($t_0$ is defined to be 0). 
    \end{itemize}
    Observe that for $i \in \{1,2,\dots,m\}$, $\clockval_{i-1} + (t_i - t_{i-1})$ is the clock valuation at time $t_i$ in {\exec} $\run$ and $\cons_i$ is the clock constraint in state $e_i$, and thus, $\clockval_i \models \cons_i$. Therefore, the {\exec} 
    $$\ptrun = (s_0, \clockval_0) \xrightarrow{e_1,\evtime_1}(s_1,\clockval_1) \xrightarrow{e_2,\evtime_2}\dots \xrightarrow{e_n,\evtime_n}(s_n,\clockval_n)$$
    is an accepting {\exec} of $\ptta$ and satisfies $\enc(\ptrun) = \run$.
    \end{proof}
    \else

    \fi
    Just as timed automata in the interval-based semantics can simulate timed automata in the point-based semantics, {\hcmtl} in the interval-based semantics is at least as expressive as {\hcmtl} in the point-based semantics. To show this, given an \hcmtl{} formula $\hcmtlf$ in the point-based semantics, we will construct a formula $\cphi{\hcmtlf}$ which when interpreted in the interval-based semantics, expresses exactly the set of all path environments that satisfy $\hcmtlf$ under a suitable encoding of path environments. Given a timed automaton $\ptta$ in the point-based semantics, and a path environment $\ptpenv: \pvar \rightarrow \ptruns{\ptta}$, we can lift the map $\chi$ to $\ptpenv$ to get a path environment $\penv_{\ptpenv}: \pvar \rightarrow \flows{\taenc{\ptta}}$ defined as $\penv_{\ptpenv}(\pi) = \enc(\ptpenv(\pi))$. Similarly, for any path environment $\penv: \pvar \rightarrow \flows{\taenc{\ptta}}$, we get a path environment $\ptpenv_{\penv}: \pvar \rightarrow \ptruns{\ptta}$ by lifting the map $\enc^{-1}$. We have the following expressiveness result:
    \begin{lemma}
    \label{pt-to-int-red}
        Given an \hcmtl{} formula $\hcmtlf$ in the point-based semantics over a set of propositions $\props$ with free path variables $\pvar$, there is a formula $\cphi{\hcmtlf}$ in the interval-based semantics over $\props \cup \{\marker\}$ such that for any timed automaton $\ptta$ in the point-based semantics and path environment $\ptpenv: \pvar \rightarrow \ptruns{\ptta}$, if $\pthcmtlmodels \hcmtlf$ then $\hcmtlargta{\penv_{\ptpenv}}{t}{\dagg}{\taenc{\ptta}} \cphi{\hcmtlf}$. Conversely, for any path environment $\penv: \pvar \rightarrow \flows{\taenc{\ptta}}$, if $\hcmtlargta{\penv}{t}{\dagg}{\taenc{\ptta}} \cphi{\hcmtlf}$ then $\pthcmtlargta{\ptpenv_\penv}{t}{\dagg}{\ptta} \hcmtlf $.
    \end{lemma}
    \ifdefined\arxiv
    \begin{proof}
    $\cphi{\hcmtlf}$ is defined inductively as follows:
    \begin{itemize}
            \item $\hcmtlf = p_\pi$. In the point-based semantics, this corresponds to $p$ being true at some event in {\exec} $\pi$ at time $t$. The interval-based semantics doesn't have a notion of an event occurring at time $t$ since the system is always observed. To capture the point-based semantics, we use the proposition $\marker$ which is true only when some transition/event occurs in $\ptta$. We define $\cphi{\hcmtlf} = \marker_\pi \wedge p_\pi$.
            \item $\hcmtlf = \neg p_\pi$. As above, we define $\cphi{\hcmtlf} = \marker_\pi \wedge \neg p_\pi$.
        \end{itemize}
        Inductive Cases
        \begin{itemize}
            \item Conjunction and disjunctions have the same semantics, so we just have
            \begin{align*}
                &\cphi{(\hcmtlf_1 \vee \hcmtlf_2)} = \cphi{\hcmtlf_1} \vee \cphi{\hcmtlf_2} \\
                &\cphi{(\hcmtlf_1 \wedge \hcmtlf_2)} = \cphi{\hcmtlf_1} \wedge \cphi{\hcmtlf_2}
            \end{align*}
            \item $\hcmtlf = F_I \hcmtlf_1$. In the point-based semantics this says that at some time $t' \in t + I$, such that an event occurs at $t'$, $\hcmtlf_1$ is true. In the interval-based semantics, we again use the $\marker$ proposition to check the truth of $\hcmtlf_1$ only at time points where event occurs. We define $\cphi{\hcmtlf}$ as: 
            $$\cphi{\hcmtlf} = F_I(\vee_{\pi \in \pvar} \marker_\pi \wedge \cphi{\hcmtlf_1}).$$
            \item $\hcmtlf = G_I \hcmtlf_1$. Similar to the case of $F_I$, we define $\cphi{\hcmtlf}$ as:
            $$\cphi{\hcmtlf} = G_I((\vee_{\pi \in \pvar} \marker_\pi) \implies \cphi{\hcmtlf_1}). $$
            \item $\hcmtlf = \hcmtlf_1 U_I \hcmtlf_2$. Using the same idea as $F_I$ we have 
            \begin{align*}
                \cphi{\hcmtlf} = (\vee_{\pi \in \pvar} \marker_\pi \implies \cphi{\hcmtlf_1})U_I (\vee_{\pi \in \pvar} \marker_\pi \wedge \cphi{\hcmtlf_2})
            \end{align*}
            \item Quantification. The semantics of existential and universal quantification is identical in both point-based and interval-based semantics. Hence, we have, 
            \begin{align*}
                &\cphi{(\exists \pi\, \hcmtlf_1)} = \exists \pi \,\cphi{\hcmtlf_1} \\
                &\cphi{(\forall \pi\, \hcmtlf_1)} = \forall  \pi\, \cphi{\hcmtlf_1}
            \end{align*}
        \end{itemize}
        We will prove correctness by inducting on the structure of $\hcmtlf$. Base Cases
        \begin{enumerate}
            \item $\hcmtlf = p_\pi$. By point-based semantics, $\hcmtlargta{\ptpenv}{t}{\dagg}{\ptta} p_\pi$ iff $t \in \ptpenv(\pi)$ and $p \in \event_{\ptpenv(\pi)}(t)$ i.e. some transition  $e = (s_1, \sigma, \cons, \gamma, s_2) \in E$ occurs at $t$ in $\ptpenv(\pi)$ and $p \in \sigma$. By definition of $\enc$, there is a singular interval $(e, [t,t])$ in $\enc(\ptpenv(\pi))$ and $p \in \stlab(e)$. Thus, $\marker_\pi \in \stlab(\penv_\ptpenv(\pi)(t))$ and $p_\pi \in \stlab(\penv_\ptpenv(\pi)(t))$, and this happens iff $\hcmtlargta{\penv_\ptpenv}{t}{\dagg}{\taenc{\ptta}} \cphi{\hcmtlf}$.
            \item The arguments for $\hcmtlf = \neg p_\pi$ are similar to the previous case.
        \end{enumerate}
        Inductive Cases:
        \begin{enumerate}
            \item $\hcmtlf = \hcmtlf_1 \vee \hcmtlf_2$. $\hcmtlargta{\ptpenv}{t}{\dagg}{\ptta} \hcmtlf_1 \vee \hcmtlf_2$ iff $\hcmtlargta{\ptpenv}{t}{\dagg}{\ptta} \hcmtlf_1$ or $\hcmtlargta{\ptpenv}{t}{\dagg}{\ptta} \hcmtlf_2$. By induction hypothesis, this happens iff $\hcmtlargta{\penv_\ptpenv}{t}{\dagg}{\taenc{\ptta}} \cphi{\hcmtlf_1}$ or $\hcmtlargta{\penv_\ptpenv}{t}{\dagg}{\taenc{\ptta}} \cphi{\hcmtlf_2}$ iff $\hcmtlargta{\penv_\ptpenv}{t}{\dagg}{\taenc{\ptta}} \cphi{\hcmtlf}$.
            \item The arguments for the conjunction is similar to the previous case.
            \item The proof of $F_I$ and $G_I$ are similar to $U_I$ which we present below.
            \item $\hcmtlf = \hcmtlf_1 U_I \hcmtlf_2$ and 
            \begin{align*}
                \cphi{\hcmtlf} = (\vee_{\pi \in \pvar} \marker_\pi \implies \cphi{\hcmtlf_1})U_I (\vee_{\pi \in \pvar} \marker_\pi \wedge \cphi{\hcmtlf_2})
            \end{align*}
            By the point-based semantics, $\hcmtlargta{\ptpenv}{t}{\dagg}{\ptta} \hcmtlf_1 U_I \hcmtlf_2$ iff there is a $t' > t$ such that $t' - t \in I$ and some event occurs at $t'$, $\hcmtlargta{\ptpenv}{t'}{\dagg}{\ptta} \hcmtlf_2$ and $\hcmtlargta{\ptpenv}{t''}{\dagg}{\ptta}\hcmtlf_1$ for all $t < t'' < t'$ such that an event occurs at $t''$. By the induction hypothesis, this happens iff 
            \begin{itemize}
                \item there is a $t' > t$ such that $t' - t \in I$
                \item $\hcmtlargta{\penv_\ptpenv}{t'}{\dagg}{\taenc{\ptta}} \vee_{\pi \in \pvar} \marker_\pi$\\
                \item $\hcmtlargta{\penv_\ptpenv}{t'}{\dagg}{\taenc{\ptta}} \cphi{\hcmtlf_2}$
                \item For all $t < t'' < t'$,
                $$\hcmtlargta{\penv_\ptpenv}{t''}{\dagg}{\taenc{\ptta}} \vee_{\pi \in \pvar} \marker_\pi \implies \cphi{\hcmtlf}.$$
                To argue this, if $\hcmtlargta{\penv_\ptpenv}{t''}{\dagg}{\taenc{\ptta}} \vee_{\pi \in \pvar} \marker_\pi$, then, by construction of $\taenc{\ptta}$ some event occurs at $t''$. By semantics of until in point-based semantics, $\hcmtlargta{\ptpenv}{t''}{\dagg}{\ptta}\hcmtlf_1$, and thus, by induction hypothesis, $\hcmtlargta{\penv_\ptpenv}{t''}{\dagg}{\taenc{\ptta}} \cphi{\hcmtlf_1}$.  
            \end{itemize}
            Hence, by semantics of until in interval-based semantics, $\hcmtlargta{\penv_\ptpenv}{t}{\dagg}{\taenc{\ptta}} \cphi{\hcmtlf}$.
            \item $\hcmtlf = \exists \pi \hcmtlf_1$. If $\dagg \in \pvar$, $\hcmtlargta{\ptpenv}{t}{\dagg}{\ptta} \exists \pi \hcmtlf_1$ iff $t \leq \len{\ptpenv(\dagg)}$ and there is an {\exec} $\ptrun \in \ptruns{\ptta}$ that extends $\dagg$ from $t$, such that for the path environment $\ptpenv'$ defined as $\ptpenv'(\pi) = \ptrun$ and $\ptpenv'(\pi') = \ptpenv(\pi')$ for $\pi'\neq \pi$, $\hcmtlargta{\ptpenv'}{t}{\dagg}{\ptta} \hcmtlf_1$. By induction hypothesis, $\hcmtlargta{\penv_{\ptpenv'}}{t}{\dagg}{\taenc{\ptta}} \cphi{\hcmtlf_1}$, and by lemma \ref{lem-bijec}, $\enc(\ptrun)(t') = \penv_{\ptpenv'}(\pi)(t')$ for all $0 \leq t' \leq t$. Thus, by semantics of existential quantifier in interval-based semantics, $\hcmtlargta{\penv_\ptpenv}{t}{\dagg}{\taenc{\ptta}} \exists \pi \cphi{\hcmtlf_1}$. 

            If $\dagg = \epsilon$, then $\run \in \ptruns{\ptta}$ has no other restrictions and the same arguments work.
            \item The argument for universal quantification is similar to the previous case.
            \end{enumerate}
            \end{proof}
    \else
    We present the construction of $\cphi{\hcmtlf}$. $\cphi{\hcmtlf}$ is defined inductively as follows:
    \begin{itemize}
            \item $\hcmtlf = p_\pi$. In the point-based semantics, this corresponds to $p$ being true at some event in {\exec} $\pi$ at time $t$. The interval-based semantics doesn't have a notion of an event occurring at time $t$ since the system is always observed. To capture the point-based semantics, we use the proposition $\marker$ which is true only when some transition/event occurs in $\ptta$. We define $\cphi{\hcmtlf} = \marker_\pi \wedge p_\pi$.
            \item $\hcmtlf = \neg p_\pi$. As above, we define $\cphi{\hcmtlf} = \marker_\pi \wedge \neg p_\pi$.
        \end{itemize}
        Inductive Cases
        \begin{itemize}
            \item Conjunction and disjunctions have the same semantics, so we just have
            \begin{align*}
                &\cphi{(\hcmtlf_1 \vee \hcmtlf_2)} = \cphi{\hcmtlf_1} \vee \cphi{\hcmtlf_2} \\
                &\cphi{(\hcmtlf_1 \wedge \hcmtlf_2)} = \cphi{\hcmtlf_1} \wedge \cphi{\hcmtlf_2}
            \end{align*}
            \item $\hcmtlf = F_I \hcmtlf_1$. In the point-based semantics this says that at some time $t' \in t + I$, such that an event occurs at $t'$, $\hcmtlf_1$ is true. In the interval-based semantics, we again use the $\marker$ proposition to check the truth of $\hcmtlf_1$ only at time points where event occurs. We define $\cphi{\hcmtlf}$ as: 
            $$\cphi{\hcmtlf} = F_I(\vee_{\pi \in \pvar} \marker_\pi \wedge \cphi{\hcmtlf_1}).$$
            \item $\hcmtlf = G_I \hcmtlf_1$. Similar to the case of $F_I$, we define $\cphi{\hcmtlf}$ as:
            $$\cphi{\hcmtlf} = G_I((\vee_{\pi \in \pvar} \marker_\pi) \implies \cphi{\hcmtlf_1}). $$
            \item $\hcmtlf = \hcmtlf_1 U_I \hcmtlf_2$. Using the same idea as $F_I$ we have 
            \begin{align*}
                \cphi{\hcmtlf} = (\vee_{\pi \in \pvar} \marker_\pi \implies \cphi{\hcmtlf_1})U_I (\vee_{\pi \in \pvar} \marker_\pi \wedge \cphi{\hcmtlf_2})
            \end{align*}
            \item Quantification. The semantics of existential and universal quantification is identical in both point-based and interval-based semantics. Hence, we have, 
            \begin{align*}
                &\cphi{(\exists \pi\, \hcmtlf_1)} = \exists \pi \,\cphi{\hcmtlf_1} \\
                &\cphi{(\forall \pi\, \hcmtlf_1)} = \forall  \pi\, \cphi{\hcmtlf_1}
            \end{align*}
        \end{itemize}
    \fi
    This translation works even if we consider the bounded time semantics by restricting all {\exec}s to a time bound $N \in \posreals$. As a corollary of this, we get
    \begin{corollary}
        \label{cor-pt-red-int}
        For any \hcmtl{} formula $\hcmtlf$ and a timed automaton $\ptta$ in the point-wise semantics,
        \begin{align*}
            \pthcmtlargta{\emptypenv}{0}{\epsilon}{\ptta} \hcmtlf \iff \hcmtlargta{\emptypenv}{0}{\epsilon}{\taenc{\ptta}} \cphi{\hcmtlf}.
        \end{align*}
    \end{corollary}
    Hence, the interval-based semantics of \hcmtl{} is at least as expressive as the point-based semantics. Thus, bounded time verification problem in point-based semantics reduces to bounded time verification problem in the interval-based semantics which gives us the following result.
    \begin{theorem}
        Bounded time verification problem for \hcmtl{} in the point-based semantics is decidable.
    \end{theorem}
    It was shown by Hsi-Ming Ho et. al.~\cite{timedhyp} that model checking HyperMITL in the point-based semantics is undecidable by reducing the universality problem of timed automata to model checking HyperMITL, which is known to be undecidable. In a similar manner, universality problem for timed automata can be reduced to the general verification problem of the \hypmitl{} fragment of \hcmtl{} in the point-based semantics. Thus, \hypmitl{} and \hcmtl{} are both undecidable in the point-based semantics. By our corollary \ref{cor-pt-red-int}, the general verification problem for \hypmitl{} and \hcmtl{} is undecidable in the interval-based semantics also. Hence, we have the following result which also implies Theorem~\ref{thm:undec}:

    \begin{theorem}
        The general verification problem for \hcmtl{} is undecidable in both the point-based and interval-based semantics. In fact, the verification problem is undecidable even for the fragment \hypmitl{}.
        \end{theorem}

\section{Related Work}
\label{sec:related}
After Clarkson and Schneider introduced hyperproperties, there has been an increasing interest in verifying hyperproperties. 
Clarkson et al. proposed temporal logic {\hypltl} and {\hypctl} to describe hyperproperties and showed that when restricted to finite Kripke structures, the model-checking problem of HyperLTL and HyperCTL* is decidable. They also establish complexity results for the verification problem.
Automated tools like MCHyper\cite{modelchecking-finkbeiner}, AutoHyper\cite{autoHyper} for model checking and satisfiability checking, EAHyper\cite{EAHyper}    for satisfiability checking and RVHyper\cite{rvhyp-finkbeiner} for runtime monitoring have been built.

To express security hyperproperties in a timed setting, different hyper-timed logics have been proposed. 
A pioneering work is \cite{hyperstl}, which extends STL with quantification over real-time signals and is studied over cyber-physical systems.


Linear-time {\sf HyperMTL} that extends {\mtl} was introduced in~\cite{borzoo}. The real-time systems in~\cite{borzoo} are modeled as timed Kripke structures, that are Kripke structures with time elapsing on transitions. The semantics used in \cite{borzoo} is point-based, and the model of time is discrete-time. The logic is defined over finite timed words but is syntactically different from ours. They show that the verification problem for the logic is decidable for a nontrivial fragment of the logic by reducing the problem to checking untimed hyperproperties. 

In \cite{khaza}, linear-time temporal logic Time Window Temporal Logic ({\sf TWTL}) is extended to reason about hyperproperties. Like \cite{borzoo}, the models are timed Kripke structures, the semantics considered is point-based, and the model of time is discrete time. The verification problem for the resulting logic is shown to be undecidable, and the model-checking algorithms are given for the alternation-free fragment of the logic.

Linear-time {\hypmitl} has been proposed in \cite{timedhyp} by extending  {\mitl}. Unlike our work, they also consider past operators in the logic. The semantics used in \cite{timedhyp} is point-based, and the model of time used is continuous/dense. \cite{timedhyp} consider decidability for many fragments of {\hypmitl}, and most of the fragments are undecidable over the unbounded time domain. The most interesting decidability result is verifying 
{\hypmitl} is decidable for bounded time domains. While the proof is not provided, the authors hint that the proof could be accomplished by a reduction to the satisfiability problem for {\QPTL}~\cite{qptl-sistla} which is known to be decidable~\cite{qptl-sistla}.

Our work extends {\mtl} to {\hcmtl} that allows us to express branching hyperproperties in addition to linear-time hyperproperties. Like~\cite{timedhyp}, the model of time is taken to be dense. However, we also consider the interval-based semantics in addition to point-based semantics. We show that the time-bounded verification for {\hcmtl} is decidable. In contrast to \cite{timedhyp}, our decidability result is obtained by a reduction to the satisfiability problem of {\sf MSO} with order and successor. The challenge is to encode the semantics of branching logic as well as time automaton into an MSO formula. We also show that the decision problems for verifying {\hcmtl} under point-based semantics can be reduced to the verification problem for interval-based semantics. This allows us to transfer decidability results under interval-based semantics to point-based semantics, and undecidability results from point-based semantics to interval-based semantics. One difference from \cite{borzoo,timedhyp} is that they distinguish between synchronous and asynchronous hyperproperties. Intuitively, in synchronous semantics, observations on all traces happen at the same time, while in asynchronous semantics they may happen at different times. Since our focus is interval-based semantics, our semantics is asynchronous in principle.

{\hyptidyctl} logic is presented in \cite{gilles1}  to reason about timed hyperproperties of timed cryptographic protocols in the Dolev-Yao model~\cite{dolev81security}. {\hyptidyctl} is also a branching logic and is similar to {\hcmtl}, but has significant differences. First, it is interpreted over timed processes and not timed automata, a variant of applied pi calculus~\cite{AbadiFournet2001} augmented with timing constructs. The logic has special
atomic constructors and atomic formulas to model the attacker's knowledge and actions. For example, it has the atomic formula $X \vdash u$, where $X$ is a variable that ranges over \emph{recipes}. Intuitively, a \emph{recipe} is a term in applied pi calculus~\cite{AbadiFournet2001} that represents a computation by a Dolev-Yao attacker from the messages it possesses, and the formula $X \vdash u$ represents that the message $u$ can be computed from the recipe assigned to $X$. The logic allows quantification over recipe and message variables in addition to the path variables.  Further, the temporal operators are not annotated explicitly with intervals, and the semantics is equivalent to taking all intervals to be $(0,\infty).$  It is shown in ~\cite{gilles1} that the problem of checking that a process satisfies a {\hyptidyctl} is undecidable, even for the {\ltl} fragment.  The proof relies on the fact that the problem of checking whether the attacker can compute a message $u$ from a finite set of messages is undecidable~\cite{AbadiCortierTCS06}.  
This undecidability result is incomparable to the undecidability of {\hcmtl} presented in this paper as both the models and the logic in~\cite{gilles1} are much richer than ours.

In ~\cite{gilles2}, a variant of {\hyptidyctl} is considered. The formula 
$X \vdash u$ is replaced by $K(u)$, which intuitively means that $u$ can be computed by the attacker from its intercepted messages using some recipe. Thus, there are no recipe variables. The quantification of message variables also takes a restricted form.  The temporal operators are now decorated with intervals as they are needed to specify liveness properties such as timeliness and fairness properties as discussed in Section~\ref{sec:examples}.    
They consider the problem of verifying whether a process satisfies a {\hyptidyctl} formula is satisfied by a process when the number of protocol sessions is bounded and when the cryptographic primitives are modeled using subterm-convergent equational theories~\cite{AbadiCortierTCS06}. They show that the verification problem becomes {\bf EXPSPACE} complete for this variant. The decision procedure is based on constraint solving. Please note that this decision procedure is incomparable to ours as we consider abstract finite timed automata and not processes. Further, the assumption of a bounded number of sessions means that the  {\lq\lq}transition system{\rq\rq} underlying a process in~\cite{gilles2} is acyclic, and any trace necessarily has a bounded number of actions/observations that depend on the process being verified. We make no such assumptions, and traces can have any number of observations.  However,    the transition system underlying ~\cite{gilles2} is infinite branching; hence, the total number of traces is still infinite in this setting. In contrast, our transition systems are finite-branching, and the number of traces is infinite because we allow for any number of actions along a trace.  

\section{conclusions and future work}
\label{sec:conclusions}

We introduce an extension of {\mtl} that expresses branching hyperproperties of real time systems. We investigate the verification problems associated with this logic against timed automata for both interval-based and point-based semantics. We show that the problem is undecidable for both semantics when the time domain is unbounded. However, when bounded time domains are considered, the verification problem becomes decidable for both semantics. 

\myparagraph{Complexity}
The decidability result is established by reducing the problem to checking the satisfaction problem of MSO with ordering and successor over bounded time domains. While this reduction establishes decidability, the complexity of deciding $\mso$ over bounded time is non-elementary\cite{boundedtime}. Hence, the decision procedure presented in this paper has non-elementary complexity. By non-elementary, we mean that the runtime cannot be bounded by a tower of exponentials whose height is independent of the automaton's size and the formula's size.  
Thus, our analysis might not provide a tight complexity bound. Given that the (space) complexity of verifying {\hypltl} and {\hypctl} for untimed systems is a tower of exponentials whose height is the alternation depth~\cite{templogichyperprop-clarkson,modelchecking-finkbeiner}, it is unlikely that {\hcmtl} will have elementary complexity over bounded time horizon.

Thus, we plan to explore the complexity of verification in terms of alternation depth. 
One potential method of obtaining a tight complexity bound is first studying the complexity of verification over rational time (which is still a dense time domain). Over bounded rational time, $\mso$ can be encoded into $\msotrees$, the monadic second order logic of two successors interpreted over binary trees. Hence, the bounded verification problem of \hcmtl{} over rational time can be reduced to a satisfiability problem over $\msotrees$. Satisfiability of $\msotrees$ has a decision procedure via a reduction to the emptiness problem of alternating tree automata. This procedure has the complexity of a tower of exponentials having a height in the order of alternation depth in the $\mso$ formula. This implies that the  complexity of bounded-time verification of \hcmtl{} when the model of dense time is the set of rationals is a tower of exponentials having height linear  in 
alternation depth in the \hcmtl{} formula. Please note that this is an upper bound on the complexity of the verification problem, and we plan to investigate the exact characterization of the complexity of verifying {\hcmtl} over a bounded time domain in the future.

\myparagraph{Algorithmic implications} 
We plan to implement the decision procedure presented in this paper using existing tools that implement decision procedures on monadic second-order logic, such as the MONA tool \cite{MONA}.

\section*{Acknowledgements}
Rohit Chadha was partially supported by NSF CCF grant 1900924, and Nabarun Deka, Minjian Zhang and Mahesh Viswanathan were partially supported by NSF SHF 1901069 and NSF CCF 2007428. We thank the anonymous reviewers for carefully reviewing our paper and providing insightful feedback and comments. 

\bibliographystyle{IEEEtranS}
\bibliography{Arxiv_TimedHyperproperties/references}

\begin{thebibliography}{10}
\providecommand{\url}[1]{#1}
\csname url@samestyle\endcsname
\providecommand{\newblock}{\relax}
\providecommand{\bibinfo}[2]{#2}
\providecommand{\BIBentrySTDinterwordspacing}{\spaceskip=0pt\relax}
\providecommand{\BIBentryALTinterwordstretchfactor}{4}
\providecommand{\BIBentryALTinterwordspacing}{\spaceskip=\fontdimen2\font plus
\BIBentryALTinterwordstretchfactor\fontdimen3\font minus \fontdimen4\font\relax}
\providecommand{\BIBforeignlanguage}[2]{{%
\expandafter\ifx\csname l@#1\endcsname\relax
\typeout{** WARNING: IEEEtranS.bst: No hyphenation pattern has been}%
\typeout{** loaded for the language `#1'. Using the pattern for}%
\typeout{** the default language instead.}%
\else
\language=\csname l@#1\endcsname
\fi
#2}}
\providecommand{\BIBdecl}{\relax}
\BIBdecl

\bibitem{AbadiCortierTCS06}
M.~Abadi and V.~Cortier, ``Deciding knowledge in security protocols under equational theories,'' \emph{Theoretical Computer Science}, vol. 387, no. 1-2, pp. 2--32, 2006.

\bibitem{AbadiFournet2001}
M.~Abadi and C.~Fournet, ``Mobile values, new names, and secure communication,'' in \emph{28th ACM Symp. on Principles of Programming Languages (POPL'01)}, 2001, pp. 104--115.

\bibitem{realtime-Alur}
\BIBentryALTinterwordspacing
R.~Alur and T.~Henzinger, ``Real-time logics: Complexity and expressiveness,'' \emph{Information and Computation}, vol. 104, no.~1, pp. 35--77, 1993. [Online]. Available: \url{https://www.sciencedirect.com/science/article/pii/S0890540183710254}
\BIBentrySTDinterwordspacing

\bibitem{alurdill}
\BIBentryALTinterwordspacing
R.~Alur and D.~L. Dill, ``A theory of timed automata,'' \emph{Theoretical Computer Science}, vol. 126, no.~2, pp. 183--235, 1994. [Online]. Available: \url{https://www.sciencedirect.com/science/article/pii/0304397594900108}
\BIBentrySTDinterwordspacing

\bibitem{mitl}
\BIBentryALTinterwordspacing
R.~Alur, T.~Feder, and T.~A. Henzinger, ``The benefits of relaxing punctuality,'' \emph{J. ACM}, vol.~43, no.~1, p. 116–146, jan 1996. [Online]. Available: \url{https://doi.org/10.1145/227595.227602}
\BIBentrySTDinterwordspacing

\bibitem{TPTL}
\BIBentryALTinterwordspacing
R.~Alur and T.~A. Henzinger, ``A really temporal logic,'' \emph{J. ACM}, vol.~41, no.~1, p. 181–203, jan 1994. [Online]. Available: \url{https://doi.org/10.1145/174644.174651}
\BIBentrySTDinterwordspacing

\bibitem{gilles1}
\BIBentryALTinterwordspacing
G.~Barthe, U.~Dal~Lago, G.~Malavolta, and I.~Rakotonirina, ``Tidy: Symbolic verification of timed cryptographic protocols,'' in \emph{Proceedings of the 2022 ACM SIGSAC Conference on Computer and Communications Security}, ser. CCS '22.\hskip 1em plus 0.5em minus 0.4em\relax New York, NY, USA: Association for Computing Machinery, 2022, p. 263–276. [Online]. Available: \url{https://doi.org/10.1145/3548606.3559343}
\BIBentrySTDinterwordspacing

\bibitem{autoHyper}
R.~Beutner and B.~Finkbeiner, ``Autohyper: Explicit-state model checking for hyperltl,'' in \emph{Tools and Algorithms for the Construction and Analysis of Systems}, S.~Sankaranarayanan and N.~Sharygina, Eds.\hskip 1em plus 0.5em minus 0.4em\relax Cham: Springer Nature Switzerland, 2023, pp. 145--163.

\bibitem{borzoo}
B.~Bonakdarpour, P.~Prabhakar, and C.~S{\'a}nchez, ``Model checking timed hyperproperties in discrete-time systems,'' in \emph{NASA Formal Methods}, R.~Lee, S.~Jha, A.~Mavridou, and D.~Giannakopoulou, Eds.\hskip 1em plus 0.5em minus 0.4em\relax Cham: Springer International Publishing, 2020, pp. 311--328.

\bibitem{khaza}
\BIBentryALTinterwordspacing
E.~Bonnah, L.~Nguyen, and K.~A. Hoque, ``Model checking time window temporal logic for hyperproperties,'' in \emph{Proceedings of the 21st ACM-IEEE International Conference on Formal Methods and Models for System Design}, ser. MEMOCODE '23.\hskip 1em plus 0.5em minus 0.4em\relax New York, NY, USA: Association for Computing Machinery, 2023, p. 100–110. [Online]. Available: \url{https://doi.org/10.1145/3610579.3611077}
\BIBentrySTDinterwordspacing

\bibitem{unlinkability}
M.~Brus{\'o}, K.~Chatzikokolakis, S.~Etalle, and J.~den Hartog, ``Linking unlinkability,'' in \emph{Trustworthy Global Computing}, C.~Palamidessi and M.~D. Ryan, Eds.\hskip 1em plus 0.5em minus 0.4em\relax Berlin, Heidelberg: Springer Berlin Heidelberg, 2013, pp. 129--144.

\bibitem{templogichyperprop-clarkson}
M.~R. Clarkson, B.~Finkbeiner, M.~Koleini, K.~K. Micinski, M.~N. Rabe, and C.~S{\'{a}}nchez, ``Temporal logics for hyperproperties,'' in \emph{Principles of Security and Trust - Third International Conference}.\hskip 1em plus 0.5em minus 0.4em\relax Springer, 2014, pp. 265--284.

\bibitem{hyperprop-clarkson}
M.~R. Clarkson and F.~B. Schneider, ``Hyperproperties,'' in \emph{Proceedings of the 21st {IEEE} Computer Security Foundations Symposium}.\hskip 1em plus 0.5em minus 0.4em\relax {IEEE} Computer Society, 2008, pp. 51--65.

\bibitem{dolev81security}
D.~Dolev and A.~Yao, ``On the security of public key protocols,'' in \emph{Proc. of the 22nd Symp. on Foundations of Computer Science}.\hskip 1em plus 0.5em minus 0.4em\relax {IEEE} Comp. Soc. Press, 1981, pp. 350--357.

\bibitem{ctl*-emerson}
E.~A. Emerson and J.~Y. Halpern, ``\enquote{Sometimes} and \enquote{Not Never} revisited: On branching versus linear time,'' in \emph{Conference Record of the Tenth Annual {ACM} Symposium on Principles of Programming Languages}.\hskip 1em plus 0.5em minus 0.4em\relax {ACM} Press, 1983, pp. 127--140.

\bibitem{EAHyper}
B.~Finkbeiner, C.~Hahn, and M.~Stenger, ``Eahyper: Satisfiability, implication, and equivalence checking of hyperproperties,'' in \emph{Computer Aided Verification}, R.~Majumdar and V.~Kun{\v{c}}ak, Eds.\hskip 1em plus 0.5em minus 0.4em\relax Cham: Springer International Publishing, 2017, pp. 564--570.

\bibitem{rvhyp-finkbeiner}
B.~Finkbeiner, C.~Hahn, M.~Stenger, and L.~Tentrup, ``{RVHyper}: {A} {Runtime} {Verification} {Tool} for {Temporal} {Hyperproperties},'' in \emph{Proceedings of the International Conference on Tools and Algorithms for the Construction and Analysis of Systems}, 2018, pp. 194--200.

\bibitem{modelchecking-finkbeiner}
B.~Finkbeiner, M.~N. Rabe, and C.~S{\'{a}}nchez, ``Algorithms for model checking {H}yper{LTL} and {H}yper{CTL}*,'' in \emph{Computer Aided Verification - 27th International Conference}.\hskip 1em plus 0.5em minus 0.4em\relax Springer, 2015, pp. 30--48.

\bibitem{MONA}
J.~Henriksen, J.~Jensen, M.~J{\o}rgensen, N.~Klarlund, B.~Paige, T.~Rauhe, and A.~Sandholm, ``Mona: Monadic second-order logic in practice,'' in \emph{Tools and Algorithms for the Construction and Analysis of Systems, First International Workshop, TACAS '95, LNCS 1019}, 1995.

\bibitem{interval-Henzinger}
T.~A. Henzinger, J.-F. Raskin, and P.-Y. Schobbens, ``The regular real-time languages,'' in \emph{Automata, Languages and Programming}, K.~G. Larsen, S.~Skyum, and G.~Winskel, Eds.\hskip 1em plus 0.5em minus 0.4em\relax Berlin, Heidelberg: Springer Berlin Heidelberg, 1998, pp. 580--591.

\bibitem{Henzinger}
T.~A. Henzinger, ``The temporal specification and verification of real-time systems,'' Ph.D. dissertation, Stanford, CA, USA, 1992, uMI Order No. GAX92-06781.

\bibitem{timedhyp}
H.-M. Ho, R.~Zhou, and T.~M. Jones, ``Timed hyperproperties,'' \emph{Information and Computation}, vol. 280, p. 104639, 2021.

\bibitem{mtl}
R.~Koymans, ``Specifying real-time properties with metric temporal logic,'' \emph{Real-time Systems}, vol.~2, no.~4, pp. 255--299, 1990.

\bibitem{obs-mclean}
J.~McLean, ``Proving noninterference and functional correctness using traces,'' \emph{Journal of Computer Security}, vol.~1, no.~1, pp. 37--58, 1992.

\bibitem{hyperstl}
\BIBentryALTinterwordspacing
L.~V. Nguyen, J.~Kapinski, X.~Jin, J.~V. Deshmukh, and T.~T. Johnson, ``Hyperproperties of real-valued signals,'' ser. MEMOCODE '17.\hskip 1em plus 0.5em minus 0.4em\relax New York, NY, USA: Association for Computing Machinery, 2017, p. 104–113. [Online]. Available: \url{https://doi.org/10.1145/3127041.3127058}
\BIBentrySTDinterwordspacing

\bibitem{WorrelMTL}
J.~Ouaknine and J.~Worrell, ``On the decidability of metric temporal logic,'' in \emph{20th Annual IEEE Symposium on Logic in Computer Science (LICS' 05)}, 2005, pp. 188--197.

\bibitem{boundedtime}
J.~Ouaknine, A.~Rabinovich, and J.~Worrell, ``Time-bounded verification,'' in \emph{CONCUR 2009 - Concurrency Theory}, M.~Bravetti and G.~Zavattaro, Eds.\hskip 1em plus 0.5em minus 0.4em\relax Berlin, Heidelberg: Springer Berlin Heidelberg, 2009, pp. 496--510.

\bibitem{worrelltalk}
J.~Ouaknine and J.~Worrell, ``Towards a theory of time-bounded verification,'' in \emph{Automata, Languages and Programming}, S.~Abramsky, C.~Gavoille, C.~Kirchner, F.~Meyer auf~der Heide, and P.~G. Spirakis, Eds.\hskip 1em plus 0.5em minus 0.4em\relax Berlin, Heidelberg: Springer Berlin Heidelberg, 2010, pp. 22--37.

\bibitem{ltl-pnueli}
A.~Pnueli, ``The temporal logic of programs,'' in \emph{Proceedings of the 18th Annual Symposium on Foundations of Computer Science}.\hskip 1em plus 0.5em minus 0.4em\relax {IEEE} Computer Society, 1977, pp. 46--57.

\bibitem{gilles2}
\BIBentryALTinterwordspacing
I.~Rakotonirina, G.~Barthe, and C.~Schneidewind, ``Decision and complexity of {D}olev-{Y}ao hyperproperties,'' \emph{Proc. ACM Program. Lang.}, vol.~8, no. POPL, jan 2024. [Online]. Available: \url{https://doi.org/10.1145/3632906}
\BIBentrySTDinterwordspacing

\bibitem{stateclock}
J.-F. Raskin and P.-Y. Schobbens, ``State clock logic: A decidable real-time logic,'' in \emph{Hybrid and Real-Time Systems}, O.~Maler, Ed.\hskip 1em plus 0.5em minus 0.4em\relax Berlin, Heidelberg: Springer Berlin Heidelberg, 1997, pp. 33--47.

\bibitem{qptl-sistla}
A.~P. Sistla, M.~Y. Vardi, and P.~Wolper, ``The complementation problem for b{\"{u}}chi automata with appplications to temporal logic,'' \emph{Theoretical Computer Science}, vol.~49, pp. 217--237, 1987.

\end{thebibliography}

\end{document}